\documentclass[aps,prxquantum, twocolumn,superscriptaddress,groupedaddress]{revtex4}  % for review and submission

\usepackage{stmaryrd}
\usepackage{physics}
\usepackage{graphicx}
\usepackage{color}
\usepackage{bm,bbold}
\usepackage{bbm}
\usepackage{enumerate}
\usepackage{color}
\usepackage{xcolor}
\usepackage{float}
\usepackage{comment}
\graphicspath{figures} % Location of the graphics files
\usepackage{amsfonts, amsmath, amsthm, amssymb}

\usepackage[colorlinks,bookmarks=false,citecolor=blue,linkcolor=red,urlcolor=blue]{hyperref}
\usepackage{enumitem}

\usepackage{cleveref}
	\crefname{equation}{Eq.}{Eqs.}
	\Crefname{equation}{Equation}{Equations}
	\crefname{figure}{Figure}{Figures}
	\Crefname{figure}{Figure}{Figures}
	\crefname{table}{Table}{Tables}
	\Crefname{table}{Table}{Tables}
	\crefname{section}{Section}{Sections}
	\Crefname{section}{Section}{Sections}
	\crefname{chapter}{Chapter}{Chapters}
	\Crefname{chapter}{Chapter}{Chapters}
	\crefname{defn}{Definition}{Definitions}	
	\Crefname{defn}{Definition}{Definitions}
	\crefname{theorem}{Theorem}{Theorems}
	\Crefname{theorem}{Theorem}{Theorems}
	\crefname{lemma}{Lemma}{Lemmas}	
	\Crefname{lemma}{Lemma}{Lemmas}		
	\crefname{cor}{Corollary}{Corollaries}		
	\Crefname{cor}{Corollary}{Corollaries}	
	\crefname{appendix}{Appendix}{Appendices}		
	\Crefname{appendix}{Appendix}{Appendices}

\newcommand\LDSeq{\mathrel{\stackrel{\makebox[0pt]{\mbox{\normalfont\tiny LDS}}}{=}}}
\newcommand\LHVMeq{\mathrel{\stackrel{\makebox[0pt]{\mbox{\normalfont\tiny LHVM}}}{=}}}

\newcommand{\ie}{\textit{i.e.} }
\newcommand{\eg}{\textit{e.g. }}

\begin{document}
\title{Deriving three-outcome permutationally invariant Bell inequalities}

\author{Albert Aloy}
\affiliation{Institute for Quantum Optics and Quantum Information, Austrian Academy of Sciences, Boltzmanngasse 3, A-1090 Vienna, Austria}
\affiliation{Vienna Center for Quantum Science and Technology (VCQ), Faculty of Physics, University of Vienna, Vienna, Austria}

\author{Guillem M\"uller-Rigat}
\affiliation{ICFO-Institut de Ciencies Fotoniques, The Barcelona Institute of Science and Technology, Castelldefels (Barcelona) 08860, Spain.}

\author{Jordi Tura}
\affiliation{$\langle aQa^L \rangle$ Applied Quantum Algorithms, Universiteit Leiden}
\affiliation{Instituut-Lorentz, Universiteit Leiden, P.O. Box 9506, 2300 RA Leiden, The Netherlands}

\author{Matteo Fadel}
\email{fadelm@phys.ethz.ch}
\affiliation{Department of Physics, ETH Z\"{urich}, 8093 Z\"{urich}, Switzerland}

\begin{abstract}
We present strategies to derive Bell inequalities valid for systems composed of many three-level parties. This scenario is formalized by a Bell experiment with $N$ observers, each of which performs one out of two possible three-outcome measurements on their share of the system. 
As the complexity of the set of classical correlations prohibits its full characterization in this multipartite scenario, we consider its projection to a lower dimensional subspace spanned by permutationally invariant one- and two-body observables. This simplification allows us to formulate two complementary methods for detecting nonlocality in multipartite three-level systems, both having a complexity independent of $N$. Our work can have interesting applications in the detection of Bell correlations in paradigmatic spin-1 models, as well as in experiments with solid-state systems or atomic ensembles.
\end{abstract}

\maketitle

\begin{figure}[t]
\centering
\includegraphics[width=0.9\linewidth]{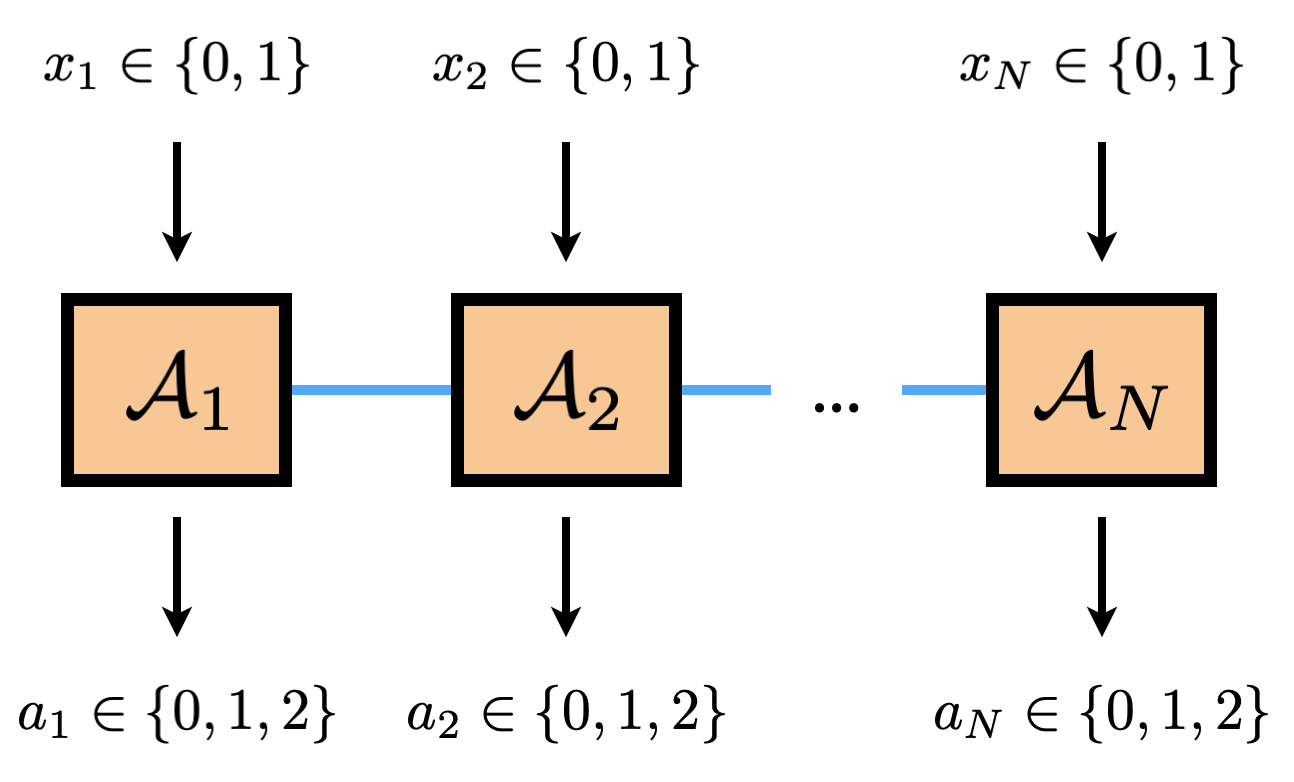}
\caption{Schematic illustration of the $(N,2,3)$ Bell scenario. Each party $\mathcal{A}_i$ performs measurement $x_i$ and observes outcome $a_i$. After many repetitions one estimates the conditional probability $p(a_1...a_N|x_1...x_N)$, to check if it is compatible with a local hidden variable description. If this is not the case, Bell nonlocality is revealed.}
\label{fig:Bell_scenario}
\end{figure}

\section{Introduction}
Bell nonlocality is a form of quantum correlation that enables some tasks inaccessible by classical means, thus constituting a key resource for quantum technologies \cite{BrunnerRMP2014,FrerotROPP2023}. Revealing the presence of nonlocality in a system is therefore of great importance, for both fundamental reasons and practical applications.

Nonlocality is typically detected by the violation of a Bell inequality, which is a criterion satisfied by any possible form of correlations with a classical explanation, such as common agreements and shared randomness. Significant effort has been put into deriving useful and practical Bell inequalities, especially in the bipartite scenario.
On the contrary, Bell inequalities valid for a large number of parties are much less explored, because characterizing the set of classical correlations becomes an intractable task even for a few parties. This can be the case already for $N=2$ parties, if the number of measurements or of possible outcomes is large \cite{BabaiCompCompl1991}.

A strategy to simplify this problem, which has proven to be successful, is to trade complexity for expressivity \cite{FrerotROPP2023}. This can be done by projecting the set of classical correlations onto a lower-dimensional subspace, for example by considering translational \cite{Tura_2014,WangPRL2017} or permutational \cite{SciencePaper,WagnerPRL2017,BaccariPRA2019,FadelPRD22,GuoPRL23} symmetries. For the multipartite scenario with two measurements and two outcomes, this approach has allowed for the derivation of Bell inequalities that enabled the study of connection between nonlocality and phase transitions \cite{FadelQuantum2018,PigaPRL2019}, metrology \cite{FroewisPRAR2019}, as well as the experimental detection of Bell correlations in spin-1/2 atomic ensembles \cite{SchmiedScience2016,EngelsenPRL2017}. Scenarios with more measurements \cite{WagnerPRL2017,GuoPRL23} or more outcomes \cite{AlsinaPRA16,GuillemPRXQuantum}, on the other hand, are much less explored.

Systems composed by spin-1 particles or, more generally, of qutrits, play an important role in nuclear physics \cite{LipkinNucPhys1965, MeshkovNucPhys1965, GlickNucPhys1965,MeredithPRA1988}, ultracold atomic ensembles \cite{LawPRL98} and in solid-state physics \cite{haldane1983continuum, HaldanePRL83, AKLTPRL87}. 
This motivated the development of ideas to simulate qudit Hamiltonians using trapped ions, superconducting circuits and ultracold atoms. 
For this reason, it is natural to ask whether Bell nonlocality can be detected in these multipartite systems, a tasks which inevitably requires many-outcome Bell inequalities.

In this work, we are interested in deriving Bell inequalities valid in a multipartite scenario with two measurements and three outcomes per party, see \cref{fig:Bell_scenario}. To make the problem tractable, we focus on finding inequalities involving permutationally invariant one- and two-body observables. Within this framework, we present two complementary strategies that allow for deriving new Bell inequalities. The first approach is based on the exact characterization of the (projected) set of classical correlations, and distilling classes of Bell inequalities that are valid for any $N$, while the second is based on a semidefinite program that approximates this set from the outside. 

Our techniques find applications in the investigation of genuinely high-dimensional nonlocality in paradigmatic three-level Hamiltonians, such as three-orbitals Lipkin-Meshkov-Glick models \cite{LipkinNucPhys1965,MeredithPRA1988,GnutzmannJPhysA1999}, as well as in the experimental detection of Bell correlations in spin-1 Bose-Einstein condensates \cite{HamleyNat12,KitzingerPRA21,OurFriends}.

\section{Bell scenario and the local polytope}

\subsection{Multipartite Bell experiment}
\label{sec:BellScenario}

We consider a Bell experiment where $N$ observers, labelled by $i \in\{1, \ldots, N\}$, perform measurements on a physical system they share. Each observer can choose to perform one out of two possible local measurements $x_i \in [m] = \{0,1,\ldots,m-1\}$ and each measurement will yield one of three possible outcomes $a_i \in [d] = \{0,1,\ldots, d-1\}$. This is known as the $(N,m,d)$ Bell scenario. In the current work, we will be interested in the $(N,2,3)$ scenario, see Fig.~\ref{fig:Bell_scenario}.

The full Bell experiment then takes place over several runs. At each run, a new copy of the physical system is distributed among the parties who will then choose a measurement to be implemented on their share and collect the resulting outcomes. Importantly, the measurement choice is independent from the state of the system and the parties are not allowed to communicate while the experiment takes place. 

After many runs, it is possible to estimate the conditional probability distribution $p(\bm{a}|\bm{x})$ for the observed statistics, which describes the probability of obtaining outcomes $\bm{a}:=(a_1,a_2,\ldots,a_N)$ when measurements $\bm{x}:=(x_1,x_2,\ldots,x_N)$ are chosen. This distribution satisfies $p(\bm{a}|\bm{x})\geq 0$ as well as the normalization condition
\begin{equation}\label{eq:pNorm}
    \sum_{\bm{a}} p(\bm{a}|\bm{x}) = 1 \qquad \forall \bm{x} \;.
\end{equation}

Moreover, since the choice of a measurement input on party $i$ cannot instantaneously signal information to the rest of the parties, the marginal probability distribution observed by any subset of the parties cannot depend on the measurements choice of the rest. 
This no-signalling (NS) principle implies the following constraint for the conditional probabilities
\begin{align}\label{eq:pNS}
 &\sum_{a_i} p(a_1, \ldots, a_i, \ldots, a_{N}| x_1, \ldots, x_i, \ldots, x_{N})  \\ 
 &\phantom{AA} =  \sum\limits_{a_i} p(a_1, \ldots, a_i, \ldots, a_{N}| x_1, \ldots, x_i^\prime, \ldots, x_{N}) \;, \nonumber
\end{align}
for all $i$ and $x_i, x_i^\prime$. Therefore, $l$-body marginals $p(a_{k_1}...a_{k_l}|x_{k_1}...x_{k_l})$ are well defined on any subset $\{k_1,...,k_l\}\subseteq \{1,...,N\}$.

% there are $(md)^N$ conditional probabilities, with $m^N$ normalization constraints

Classical correlations imply that $p(\bm{a}|\bm{x})$ can be explained by pre-established agreements, \ie by a local hidden variable model (LHVM), meaning that 
\begin{equation}\label{eq:pLHVM}
     p(\bm{a}|\bm{x}) \;\;\LHVMeq\; \int p(\lambda) p(a_1|x_1,\lambda)...p(a_N|x_N,\lambda)  \mathrm{d}\lambda \;,
\end{equation}
where $p(\lambda)$ is the probability of using agreement $\lambda$. If this is not the case, namely if the observed statistics $p(\bm{a}|\bm{x})$ cannot be written as in \cref{eq:pLHVM}, we have to conclude the presence of Bell correlations in the system, \ie nonlocality.

The set of $p(\bm{a}|\bm{x})$ for all $\bm{a}$ and $\bm{x}$ can be interpreted as a real vector in a probability space. Crucially, in this geometrical picture, the set of LHVM correlations is a polytope $\mathbb{P}$, \ie a bounded, closed, convex set with a finite number of extreme points, called vertices. $\mathbb{P}$ can be described by a such a finite number of vertices, or, equivalently, by the intersection of a finite number of half-spaces. These half-spaces are defined by linear inequalities that take the name of Bell inequalities.
Concluding the presence of Bell nonlocality thus consists in deciding (non-)membership in the local polytope $\mathbb{P}$. 
In the following sections we want to propose strategies to tackle this membership problem for the $(N,2,3)$ scenario.

\subsection{Local Deterministic Strategies and characterization of the Local Polytope}
\label{sec:LDS}

The vertices of the local polytope $\mathbb{P}$ correspond to a special class of LHVM, namely local deterministic strategies (LDS) \cite{FinePRL1982}. These satisfy
\begin{align}
& p(\bm{a}|\bm{x}) = p(a_1|x_1)...p(a_N|x_N)  \quad \text{(local)} \label{eq:local} \\
& p(a_i|x_i) = 0 \;\text{or}\; 1  \quad \forall\, i, a_i, x_i  \qquad \text{(deterministic).} \label{eq:deterministic}
\end{align}
Equation \eqref{eq:deterministic} implies that for a single party with $m$ measurement settings (inputs) and $d$ possible outcomes (outputs) there are $d^m$ LDS. For the $m=2$, $d=3$ case we consider in this work, the nine single-party LDS are illustrated in Tab.~\ref{table:N23LDS}. 

From the single-party LDS and Eq.~\eqref{eq:local}, it is in principle possible to list all $d^{mN}$ vertices of $\mathbb{P}$.
Revealing nonlocality then consists in asking whether the observed statistics $p(\bm{a}|\bm{x})$ can be written as a convex combination of these vertices, which can be expressed as a linear program (LP) for which there exist efficient numerical algorithms. However, note that the exponential scaling in the number of parties makes such an approach unfeasible for large $N$.

Alternatively, it is possible to use computational geometry algorithms to convert the list of vertices (V-representation of $\mathbb{P}$) into a list of inequality constraints (H-representation of $\mathbb{P}$). These constraints specify the facets of the local polytope, and thus corresponds to tight Bell inequalities. Revealing nonlocality then consists in asking whether (at least) one of these inequalities is violated, as it would imply the impossibility of explaining the observed statistics by a LHVM. 

In general, the number of inequalities can be much larger than the number of vertices (and, by duality, vice-versa). A paradigmatic example where this occurs is the cross-polytope
$C_d = \{\mathbf{x} \in {\mathbb R}^d:\ ||\mathbf{x}||_1\leq 1\} $, 
which has $2d$ vertices but $2^d$ inequalities. This worst-case exponential scaling manifest itself in dual-description algorithms, that run in complexity as $O(v^{\lfloor D/2 \rfloor})$, where $v$ is the number of vertices and $D$ the polytope dimension \cite{ChazelleDCG1993}.

Therefore, since the number of vertices scales exponentially with $N$, an exhaustive enumeration of all the vertices is typically impossible in a multipartite scenario. 
For example, in the $(N,2,3)$ scenario considered here, $N=10$ already gives $v \sim 10^9$ vertices, which is already a prohibitive number for commonly available computing resources.

Deriving and characterizing the polytope of classical correlations for a multipartite system thus poses a formidable challenge. Already for a few parties, the underlying combinatorial complexity \cite{BabaiCompCompl1991, PitowskyPRA2001} of the problem makes the task of listing all vertices or all facets unapproachable. For this reason, in order to partially overcome this complexity, it is often chosen to perform some simplification to the problem. This allows one to trade-off some nonlocality detection capability for the derivation of useful multipartite Bell inequalities. Moreover, we can do so in a way that the resulting inequalities fulfill desirable properties that make their experimental verification more approachable. An important strategy to simplify the complexity of the local polytope is discussed in the next section.

\begin{table}[t]
\begin{center}
 \begin{tabular}{|c | c c c c c c |} 
 \hline
 LDS label & $p(0|0)$ & $p(1|0)$ & $p(2|0)$ & $p(0|1)$ & $p(1|1)$ & $p(2|1)$\\
 \hline
 \# 0 & 1 & 0 & 0 & 1 & 0 & 0\\ 
 \# 1 & 1 & 0 & 0 &  0  & 1 & 0\\
 \# 2 & 1 & 0 & 0 &  0  & 0 & 1\\
 \# 3 & 0 & 1 & 0 & 1 & 0 & 0\\ 
 \# 4 & 0 & 1 & 0 &  0  & 1 & 0\\
 \# 5 & 0 & 1 & 0 &  0  & 0 & 1\\ 
 \# 6 & 0 & 0 & 1 & 1 & 0 & 0\\ 
 \# 7 & 0 & 0 & 1 &  0  & 1 & 0\\
 \# 8 & 0 & 0 & 1 &  0  & 0 & 1\\ 
 \hline
\end{tabular}
\end{center}
\caption{Local deterministic strategies $p(a|x)$ in the 2-input 3-output scenario. Note that they satisfy $\sum_a p(a|x)=1$.} \label{table:N23LDS}
\end{table}

\begin{table*}[t]
\renewcommand{\arraystretch}{1.7}
\centering
\begin{tabular}{| @{\hskip 0.1in}l@{\hskip 0.1in}|@{\hskip 0.1in}l@{\hskip 0.1in}|@{\hskip 0.1in}l@{\hskip 0.1in}|@{\hskip 0.1in}l@{\hskip 0.1in} |}
\hline
$\mathcal{P}_{0|0}\LDSeq c_{0,0}+c_{0,1}+c_{0,2}$ & $\mathcal{P}_{00|00}\LDSeq\mathcal{P}_{0|0}^2-\mathcal{P}_{0|0}$ & $\mathcal{P}_{00|01}\LDSeq\mathcal{P}_{0|0}\mathcal{P}_{0|1}-c_{0,0}$ & $\mathcal{P}_{00|11}\LDSeq\mathcal{P}_{0|1}^2-\mathcal{P}_{0|1}$ \\
$\mathcal{P}_{0|1}\LDSeq c_{0,0}+c_{1,0}+c_{2,0}$ & $\mathcal{P}_{01|00}\LDSeq\mathcal{P}_{0|0}\mathcal{P}_{1|0}$ & $\mathcal{P}_{01|01}\LDSeq\mathcal{P}_{0|0}\mathcal{P}_{1|1}-c_{0,1}$ & $\mathcal{P}_{01|11}\LDSeq\mathcal{P}_{0|1}\mathcal{P}_{1|1}$\\
$\mathcal{P}_{1|0}\LDSeq c_{1,0}+c_{1,1}+c_{1,2}$ & $\mathcal{P}_{11|00}\LDSeq\mathcal{P}_{1|0}^2-\mathcal{P}_{1|0}$ & $\mathcal{P}_{10|01}\LDSeq\mathcal{P}_{1|0}\mathcal{P}_{0|1}-c_{1,0}$ & $\mathcal{P}_{11|11}\LDSeq\mathcal{P}_{1|1}^2-\mathcal{P}_{1|1}$ \\
$\mathcal{P}_{1|1}\LDSeq c_{0,1}+c_{1,1}+c_{2,1}$ &  &  $\mathcal{P}_{11|01}\LDSeq\mathcal{P}_{1|0}\mathcal{P}_{1|1}-c_{1,1}$ &  \\
\hline
\end{tabular}
\caption{Value of one- and two-body PI observables when computed for a LDS. The variables $c_{a,a'}$ are non-negative integers satisfying $\sum_{a,a'}c_{a,a'}=N$, see main text.} \label{table:LDS}
\end{table*}

\subsection{Projections onto low-dimensional subspaces}
\label{sec:Proj}

\begin{figure}[t]
\centering
\includegraphics[width=0.85\columnwidth]{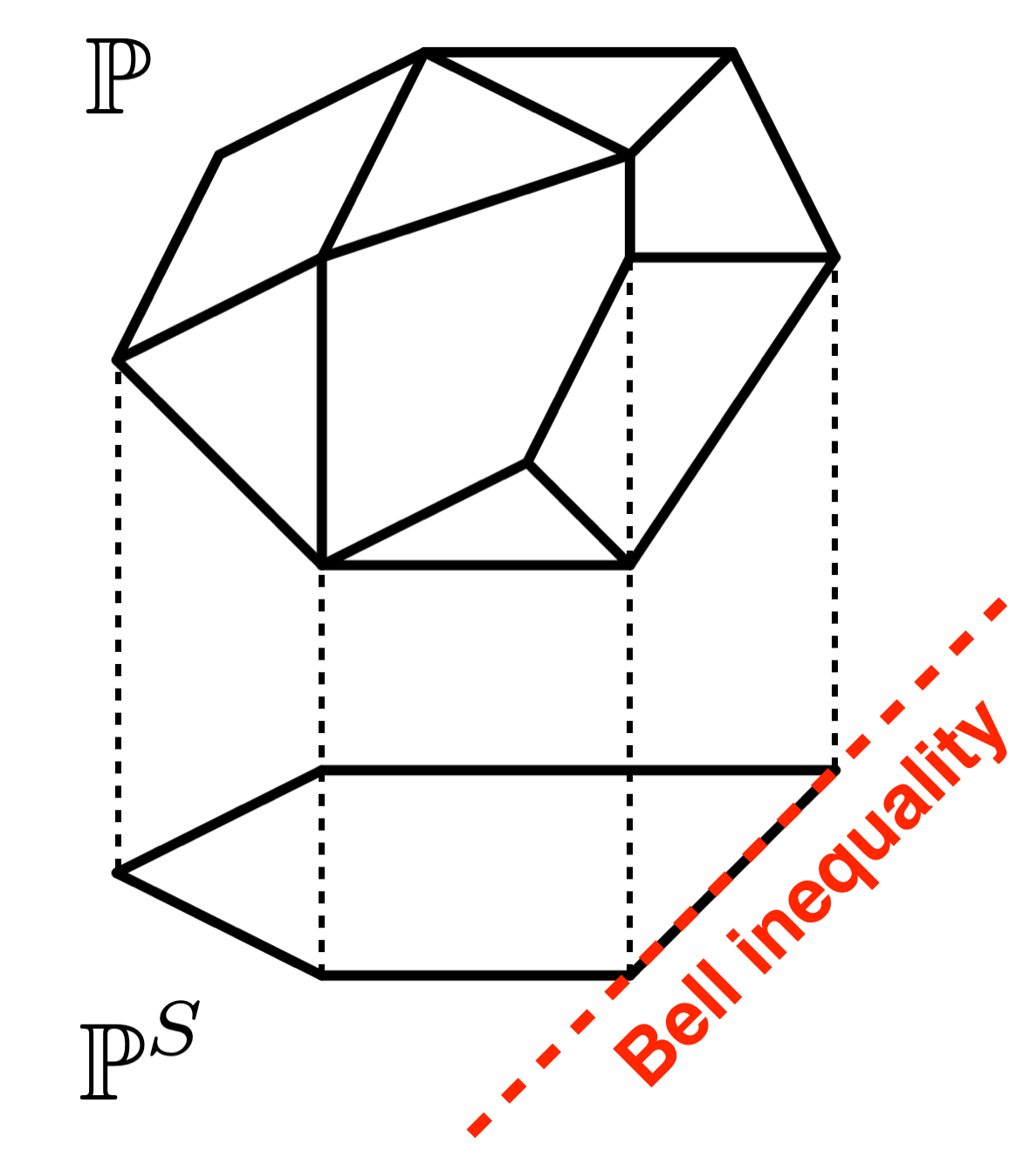}
\caption{Illustration of the local polytope $\mathbb{P}$ and of its projection $\mathbb{P}^S$ to a lower dimensional subspace. The red dashed line indicates a resulting Bell inequality in the projected space.
}
\label{fig:poly}
\end{figure}

A particularly successful approach to simplify the membership problem just mentioned consists of projecting $\mathbb{P}$ onto a lower dimensional subspace. The resulting projection is also a polytope, $\mathbb{P}^S$, of easier characterization because of the reduced dimensionality, see \Cref{fig:poly}. If the observed statistics $p(\bm{a}|\bm{x})$, after being appropriately projected, does not belong to $\mathbb{P}^S$, then we must conclude that it also does not belong to $\mathbb{P}$ and thus that nonlocality is detected. On the contrary, if the projected $p(\bm{a}|\bm{x})$ lies inside $\mathbb{P}^S$, then we cannot know whether $p(\bm{a}|\bm{x})$ lies inside $\mathbb{P}$ or not. For this reason, it is important to choose a projection that does not result in losing too much information about $\mathbb{P}$, such that nonlocality can be detected, while still reducing the complexity of $\mathbb{P}$ to a computationally manageable level.

In the following, we consider a projection to the space of permutationally invariant (PI) one- and two-body conditional probabilities, that is defined by the observables
\begin{align}
\mathcal{P}_{a|x} &:= \sum\limits_{i} p(a_i|x_i) \;, \label{eq:pPI1}\\
\mathcal{P}_{ab|xy} &:= \sum\limits_{i\neq j} p(a_i b_j|x_i y_j) \label{eq:pPI2} \;.
\end{align}

Limiting ourselves to one- and two-body marginals reduces the probability space from $[m(d-1)+1]^N-1$ dimensions (all possible $p(\bm{a}|\bm{x})$ minus the constraints Eq.~\eqref{eq:pNorm}) to $[Nm(d-1) + {N \choose 2}(m(d-1))^2]$ dimensions. Despite this being a significant simplification, characterizing the projection of the local polytope in this subspace still requires to enumerate all $d^{mN} = 3^{2N}$ vertices. For this reason, we consider PI observables, which has two important consequences.

First, the dimension of the PI probability space defined by Eqs.~(\ref{eq:pPI1},\ref{eq:pPI2}) is 
reduced to $m(d-1) + m(d-1) + {m(d-1) \choose 2} \in O(1)$ dimensions, which is independent of the number of parties $N$. Second, the vertices of the projected local polytope can now be parametrized much more efficiently: instead of having to consider all the $d^{m N}$ combinations specifying which party uses which LDS, permutational invariance implies that only knowing the number of particles using each LDS matters. Therefore, it is sufficient to consider partitions of $N$ into $d^m$ integers, resulting in a list of points of polynomial size ${N+d^m-1 \choose d^m-1}\sim O(N^{d^m-1}/(d^m-1)!)$. Some of these points will be vertices of the projected local polytope $\mathbb{P}^S$, while the rest are in general interior points. The polynomial scaling may be improved to a lower degree by studying further the structure of the projected vertices. For instance, in \cite{AnnPhys} for the $(N,2,2)$ scenario the scaling could be improved from $N^3$ to $N^2$. We expect, but do not prove, that this improvement holds in more general scenarios. This is a consequence of most of the projected vertices not being extremal in the projection.

%\al{\subsubsection{Local Deterministic Strategy for 3-outcome PIBIs}}
Having in mind the $(N,2,3)$ scenario, let us denote with $c_{a,a'}$ be the total number of parties that have predetermined the pair of outcomes $a,a'\in \{0,1,2\}$ for the two measurement settings $x=0$ and $x=1$ respectively. For example, $c_{2,0}$ is the number of parties giving as outcome $a=2$ when $x=0$ is measured and $a'=0$ when $x=1$ is measured. That is, $c_{2,0}$ corresponds to the number of parties following the LDS $\# 6$ in Tab.~\ref{table:N23LDS}. Note that $a,a'$ corresponds to the expression of the LDS label in base $d$ (so, $6 = 20_{(3)}$).

It follows by definition that the integers $c_{a,a'}\geq 0$ and $\sum_{a,a'}c_{a,a'}=N$. Adopting this parametrization, it is possible to express the value of the PI one-body observables \cref{eq:pPI1} when computed on a LDS as
\begin{align}
\mathcal{P}_{a|x} \LDSeq \left\lbrace\begin{array}{cc}
    c_{a,0}+c_{a,1}+c_{a,2} & \textrm{for }x=0 \\
    c_{0,a}+c_{1,a}+c_{2,a} & \textrm{for }x=1
\end{array}\right. \;.
\label{eq:oneProb}
\end{align}
Similarly, the PI two-body observables \cref{eq:pPI2} factorize under a given LDS as
\begin{equation} 
\begin{split}
 \mathcal{P}_{ab|xy} & \LDSeq \sum\limits_{i\neq j} p(a_i|x_i)p(b_j|y_j) \\
&= \underbrace{\sum\limits_{i\neq j}p(a_i|x_i)p(b_j|y_j)+\sum\limits_i p(a_i|x_i)p(b_i|y_i)}_{\mathcal{P}_{a|x}\cdot \mathcal{P}_{b|y}} \\
&\phantom{==} - \underbrace{\sum\limits_i p(a_i|x_i)p(b_i|y_i)}_{:=\mathcal{Q}_{ab|xy}} \;, \label{eq:twoProb}
\end{split}
\end{equation}
where we have
\begin{align}
\mathcal{Q}_{ab,xy}=
\left\{
\begin{array}{l}
\mathcal{P}_{a|x}\\
0 \\
c_{a,b}\\ 
c_{b,a}
\end{array}\right.
\begin{array}{l}
\text{if }a=b,x=y\\
\text{if }a\neq b,x=y\\ 
\text{if }x=0, y=1\\ 
\text{if }x=1, y=0 
\end{array} \;.
\label{eq:Q}
\end{align} 

We summarize in \Cref{table:LDS} the one-body terms and the factorized two-body terms as a function of the quantities $c_{a,a'}$. 
Note that, without loss of generality, we used the NS principle \cref{eq:pNS} together with the normalization condition \cref{eq:pNorm} to neglect one of the outcomes (in this work, we choose to eliminate the $(d-1)$-th outcome).
For example, because we can write
\begin{align}
    p(02|xy) &= 1 - p(00|xy)-p(01|xy)-p(10|xy)-p(11|xy) \notag\\
    &\phantom{AA} -p(12|xy) -p(20|xy)-p(21|xy)-p(22|xy) \notag\\
    &= 1 - p(00|xy)-p(01|xy)-p(1|x)-p(2|x) \notag\\
    &= 1 - p(00|xy)-p(01|xy)- (1-p(0|x) )\notag \\
    &= p(0|x) - p(00|xy)-p(01|xy)\notag \;.
\end{align}
Hence, in our scenario we can always consider Eqs.~(\ref{eq:oneProb}--\ref{eq:Q}) with \eg $a,b\in\{0,1\}$ only. 
In addition, note that PI results in redundancies of the two-body observables \cref{eq:pPI2}, such as ${\mathcal P}_{01|10}={\mathcal P}_{10|01}$ or ${\mathcal P}_{10|00}={\mathcal P}_{01|00}$. For this reason, for ${\mathcal P}_{ab|xy}$ we take $x\leq y$ as canonical notation and, when $x=y$, we choose $a \leq b$.

From the above considerations, we can conclude that the projected local polytope $\mathbb{P}^S$ for the $(N,2,3)$ scenario can be expressed in the 14-dimensional space of coordinates
\begin{align} \label{eq:Coord14dim}
\{&\mathcal{P}_{0|0}, \mathcal{P}_{1|0}, \mathcal{P}_{0|1}, \mathcal{P}_{1|1}, \mathcal{P}_{00|00}, \mathcal{P}_{01|00}, \mathcal{P}_{11|00}, 
       \\
&\; \mathcal{P}_{00|01},\mathcal{P}_{01|01}, 
     \mathcal{P}_{10|01},\mathcal{P}_{11|01}, 
     \mathcal{P}_{00|11}, \mathcal{P}_{01|11}, \mathcal{P}_{11|11} \} \;.\nonumber
\end{align}
The vertices of $\mathbb{P}^S$ can thus be found by computing \cref{eq:Coord14dim} for all possible LDS using the relations given in \cref{table:LDS}. This gives a list of ${N+3^2-1 \choose 3^2-1}\sim O(N^8/8!)$ points, some of which are vertices.
In the following, we are going to show how to use this list of points or, alternatively, the expressions in \cref{table:LDS}, to derive useful three-outcome Bell inequalities for multipartite systems.

\section{Deriving new multipartite Bell inequalities}
\label{sec:DerivationBI}

After having characterized a projection of the local polytope in terms of its vertices, we are now interested in deriving Bell inequality capable of revealing Bell correlations in multipartite systems composed of three-level constituents. Since we consider a projection onto the space of permutationally invariant one- and two-body observables Eqs.~(\ref{eq:pPI1},\ref{eq:pPI2}), we call our inequalities 3-level PI Bell inequalities (3PIBIs).
 
We start with presenting in \Cref{sec:Class} the more conventional (albeit somewhat involved) approach to derive new 3PIBIs, which consists of computing complete lists of inequalities for small $N$ to then extrapolate how they generalize to arbitrary large $N$. Then, we present in section \Cref{sec:ThetaBodiesMethod} a data-driven method to derive new inequality, which consists in using a numerical algorithm that approximates the local polytope from the outside to check membership for the experimentally observed statistics. The second method can be seen as checking all the inequalities defining an outer approximation to $\mathbb{P}^S$ at once.

\subsection{Inferring families of 3-outcome PIBIs}
\label{sec:Class}

We have seen in \Cref{sec:Proj} the approach of projecting the local polytope $\mathbb{P}$ for the $(N,2,3)$ scenario onto a lower dimensional subspace, which results in the polytope $\mathbb{P}^S$ of much lower complexity: it lives in a space that is 14-dimensional, independently of the number of parties $N$, and its number of vertices scales as $N^2$.  

For a given $N$, the vertices of $\mathbb{P}^S$ are obtained by evaluating \cref{eq:Coord14dim} for all LDS using \cref{table:LDS}.
%For sufficiently small values of $N$ it is thus possible to fully enumerate the vertices of $\mathbb{P}^S$ and, eventually, to obtain from them the full list its facets.
From this list of points it is in principle possible to use computer algorithms \cite{fukuda1997cdd} to obtain the dual description of $\mathbb{P}^S$ in terms of intersecting half-spaces. Each one of this half-spaces is specified by an inequality constraint, which corresponds to a valid Bell inequality that is also tight (\ie is a facet) for $\mathbb{P}^S$. 
These take the form
\begin{equation}
B=\sum\limits_{a,x}\alpha_{a,x}\mathcal{P}_{a|x}+\sum\limits_{a,b,x,y}\alpha_{ab,xy}\mathcal{P}_{ab|xy} + \beta_c \geq 0 \;, 
\label{eq:gralIneq}
\end{equation}
where the summations run over the terms in \cref{eq:Coord14dim}, $\alpha_{\bm{a},\bm{x}}\in\mathbbm{R}$ and the real number
\begin{equation} \label{betaC}
\beta_c := - \min_{\text{LDS}} \left( \sum\limits_{a,x}\alpha_{a,x}\mathcal{P}_{a|x}+\sum\limits_{a,b,x,y}\alpha_{ab,xy}\mathcal{P}_{ab|xy} \right)
\end{equation}
is the so called classical bound. 
In general, all the parameters $\alpha$ and $\beta_c$ depend on $N$. If a system exhibits statistics such that $B < 0$, then it cannot be explained by a LHVM and Bell nonlocality is detected.

%or, equivalently, the resulting affine space is a 15-dimensional vector space in the coordinates \mf{write this more precisely}
%
%\begin{align} \label{eq:15dim}
%\{&1, \mathcal{P}_{0|0}, \mathcal{P}_{0|1}, \mathcal{P}_{1|0}, \mathcal{P}_{1|1}, \mathcal{P}_{00|00}, \mathcal{P}_{10|00}, \mathcal{P}_{11|00}, 
%       \\
%&\mathcal{P}_{01|01},\mathcal{P}_{10|01}, 
%     \mathcal{P}_{00|10},\mathcal{P}_{11|10}, 
%     \mathcal{P}_{00|11}, \mathcal{P}_{10|11}, \mathcal{P}_{11|11} \}.\nonumber
%\end{align}
%Note that, in order to obtain the coordinates in \cref{eq:15dim}, we have omitted without loss of generality some terms such as ${\mathcal P}_{00|01}$ or ${\mathcal P}_{01|11}$ \mf{what about ${\mathcal P}_{01|00}$}, since they result in redundancies due to permutation invariance. Henceforth, for the terms ${\mathcal P}_{ab|xy}$ we take $x\leq y$ as canonical notation  and, when $x=y$, we choose $a \leq b$.

In our analysis, we have been able to apply this procedure for $N=2,3$, which resulted in $165$ and $146994$ 3PIBIs respectively. 
It is interesting to note here that, while the number of vertices can be computed exactly to be $\sim N^8$, the number of inequalities can have a severe scaling, in the worst case, $O(\mathrm{poly}(N)^{O(1)})$, which is still polynomial, but of a worse degree.
For $N>3$ the computational resources required by the algorithm converting vertices into facets was prohibitive, thus making impossible to pursue this numerical approach further. 
%\mf{here we have to mention why we want to extrapolate to large N} Despite the possibility to isolate from the set of inequalities we have found the ones admitting a quantum violation \mf{CITE}, it is practically unfeasible to conjecture generalizations to arbitrary $N$ by looking only at the cases $N=2,3$.

For this reason, in order to be able to find inequalities that are be valid for a large number of parties, we impose an additional simplification to the problem. We decide to look only for 3PIBIs that are also invariant under relabeling of inputs and outputs 0, 1. This choice can be motivated from physical arguments: one can imagine interactions in spin-1 systems, such as spin-exchanging collisions \cite{HamleyNat12,KitzingerPRA21,OurFriends}, that result in states that are invariant under relabeling of $S_z=+1$ with $S_z=-1$. This additional symmetry we require for the 3PIBI can be enforced by introducing the following symmetrized one- and two-body observables
\begin{align}
\tilde{\mathcal{P}}_{0} &:= \mathcal{P}_{0|0}+P_{0|1} + P_{1|0}+P_{1|1} \\
\tilde{\mathcal{P}}_{00}  &:= \mathcal{P}_{00|00} + \mathcal{P}_{00|11} + \mathcal{P}_{11|00} + \mathcal{P}_{11|11}  \\
\tilde{\mathcal{P}}_{01} &:=  \mathcal{P}_{01|01} + \mathcal{P}_{10|01} \\
\tilde{\mathcal{P}}_{10}  &:=  \mathcal{P}_{00|01} + \mathcal{P}_{11|01} \\
\tilde{\mathcal{P}}_{11}  &:=  \mathcal{P}_{01|00} + \mathcal{P}_{01|11} \;.
\end{align}
The 14-dimensional space defined by \cref{eq:Coord14dim} of $\mathbb{P}^S$ is thus projected further onto the 5-dimensional space with coordinates
\begin{equation} \label{eq:6dim}
\{ \tilde{\mathcal{P}}_{0}, \tilde{\mathcal{P}}_{00},\tilde{\mathcal{P}}_{01}, \tilde{\mathcal{P}}_{10},\tilde{\mathcal{P}}_{11}\} \;. 
\end{equation}
Note that, by working in this subspace, we are effectively looking for 3PIBIs of the form
\begin{equation}\label{eq:5dimPIBI}
    B = \alpha_1 \tilde{\mathcal{P}}_{0}  + \alpha_2 \tilde{\mathcal{P}}_{00} + \alpha_3 \tilde{\mathcal{P}}_{01} + \alpha_4 \tilde{\mathcal{P}}_{10} + \alpha_5 \tilde{\mathcal{P}}_{11} + \beta_c \geq 0 \;,
\end{equation}
which correspond to instances of \cref{eq:gralIneq} with a very specific relation between the coefficients, namely $\alpha_{0,0}=\alpha_{0,1}=\alpha_{1,0}=\alpha_{1,1}$, $\alpha_{00,00}=\alpha_{00,11}=\alpha_{11,00}=\alpha_{11,11}$, $\alpha_{01,01}=\alpha_{10,01}$, $\alpha_{00,01}=\alpha_{11,01}$ and $\alpha_{01,00}=\alpha_{01,11}$. 

With this additional projection we were able to compute full list of vertices and obtain from them complete lists of inequalities up to $N= 17$, for which we find $1415$ 3PIBIs. Again, accessing even larger number of parties requires significant computational resources, thus becoming quickly unfeasible. However, having at hand complete lists of inequalities for $2 \leq N \leq 17$ allows us to look for patterns in the coefficients, or for recurrent inequalities, and conjecture possible 3PIBIs that could be valid for arbitrarily large $N$.
To give a concrete example, we propose for any $N>3$ the five 3PIBIs shown in Tab.~\ref{table:3PIBIS}.
\begin{table}[t]
\begin{center}\label{table:otherPIBIs}
 \begin{tabular}{|c | c c c c c c|} 
 \hline
 3PIBI label & $\alpha_1$ & $\alpha_2$ &$\alpha_3$ & $\alpha_4$ & $\alpha_5$ & $\beta_c$ \\
 \hline
 \# 1 & 1 & 1 & 0 & -2 & 0 & 0 \\
 \hline
 \# 2 & 1 & 1 & -2 & -2 & 2 & 0 \\ 
 \hline
 \# 3 & -2 &  1  & 2 & 2 & 0 & 4 \\
 \hline
 \# 4 & -6 &  1 &  4 & 4 & 2 & 12 \\
 \hline
 \# 5 & -6 & 1 & 4 & 0 & 0 & 24 \\ 
 \hline
\end{tabular}
\end{center}
\caption{Five proposed families of 3PIBIs, see Eq.~\eqref{eq:5dimPIBI}.} \label{table:3PIBIS}
\end{table}

At this point, for each conjectured inequality we have to prove that it is indeed valid for arbitrary number of parties $N$, or at least for all $N$ larger than a minimum number. To this end, we use again the prescription given by \Cref{table:LDS} to write each inequality as a polynomial function in the coefficients $c_{a,a'}$. Then, we want to prove that this function is nonnegative when evaluated for integer values of the $c_{a,a'}$. This ensures that the porposed 3PIBIs cannot be violated by a LHVM, and that it is thus a valid Bell inequality. A concrete example of how this is done for PIBI $\#$1 can be found in Refs.~\cite{UsInPrep,UsInPrep2}.

Finally, once a new Bell inequality is found, it is left to verify that it can be violated by appropriate measurements on a quantum state. This task is relatively easy to tackle in the $(N,m,2)$ scenario \cite{AloyQMP,WagnerPRL2017,GuoPRL23}, where one can consider $N$ qubits on which Pauli measurements are performed among $m$ directions (independent on the party).
For the $(N,2,3)$ scenario, the task of looking for a quantum violation can be significantly more tedious. In particular, parametrizing qutrit measurements from SU(3) operators is more involved. For this reason, we refer the reader interested in this search for a quantum violation to Refs.~\cite{UsInPrep,UsInPrep2}.

\subsection{Data-driven derivation of 3-outcome PIBIs}
\label{sec:ThetaBodiesMethod}

The approach to derive multipartite 3PIBIs proposed in the previous section consisted of deriving complete lists of inequalities for small $N$, to then look for inequalities that could be generalized to arbitrary large $N$. However, due to the complexity of the problem, we had to further restrict our search to 3PIBIs that are invariant under relabeling of some inputs and outputs, see \cref{eq:5dimPIBI}, therefore losing the possibility to find more general inequalities of the form \cref{eq:gralIneq}.

Here, we propose a complementary approach based on a generalization of Ref.~\cite{FadelPRL2017} for an arbitrary number of outcomes. 
The general idea consists in approximating $\mathbb{P}^S$ from the outside by the convex hull of a semialgebraic set (\ie of a set defined by polynomial equalities and inequalities), to then write a semidefinite program (SdP) whose infeasibility is sufficient to certify that the observed statistics $p(\bm{a}|\bm{x})$ lies outside $\mathbb{P}^S$ and, thus, that it exhibits nonlocal correlations. Importantly, SdPs form a class of well-behaved convex optimization problems which can be efficiently solved by numerical routines \cite{CVX1}. In addition, the dual variables of an infeasible SdP \cite{BookSDP} provide a Bell inequality which is violated by $p(\bm{a}|\bm{x})$.
Without going into the mathematical details of the method, in the following we want to describe how to write such SdP.

Our first task is to derive an outer approximation of $\mathbb{P}^S$ in terms of the convex hull of a semialgebraic set. To this end, we consider the coefficients $c_{a,a'}$ appearing in \cref{table:LDS} to be non-negative real numbers, instead of positive integers. Then, we invert the expressions in \cref{table:LDS} to write the conditions $c_{a,a'}\in\mathbb{R}_{\geq 0}$ in terms of the $\mathcal{P}_{\boldsymbol{a}|\boldsymbol{x}}$, as well as to write the algebraic constraints between them. These are
\begin{align}\label{eq:fEqs}
\left\{\begin{array}{l}
\mathcal{P}_{0|0}^2 - \mathcal{P}_{0|0} - \mathcal{P}_{00|00}  = 0\\
\mathcal{P}_{0|0}\mathcal{P}_{1|0} - \mathcal{P}_{01|00} = 0\\
\mathcal{P}_{1|0}^2 - \mathcal{P}_{1|0} - \mathcal{P}_{11|00} = 0\\
\mathcal{P}_{0|1}^2 - \mathcal{P}_{0|1} - \mathcal{P}_{00|11} = 0\\
\mathcal{P}_{1|1}\mathcal{P}_{0|1} - \mathcal{P}_{01|11} = 0\\
\mathcal{P}_{1|1}^2 - \mathcal{P}_{1|1} - \mathcal{P}_{11|11} = 0
\end{array}\right.
\end{align}
together with
\begin{widetext}
\begin{align} \label{eq:gEqs}
\left\{\begin{array}{l}
c_{0,0}=\mathcal{P}_{0|0}\mathcal{P}_{0|1}-\mathcal{P}_{00|01}\geq 0\\
c_{0,1}=\mathcal{P}_{0|0}\mathcal{P}_{1|1}-\mathcal{P}_{01|01}\geq 0\\
c_{0,2}=\mathcal{P}_{0|0}-c_{0,0}-c_{0,1}= \mathcal{P}_{0|0}-\mathcal{P}_{0|0}\mathcal{P}_{0|1}+\mathcal{P}_{00|01}-\mathcal{P}_{0|0}\mathcal{P}_{1|1}+\mathcal{P}_{01|01}\geq 0\\
c_{1,0}=\mathcal{P}_{1|0}\mathcal{P}_{0|1}-\mathcal{P}_{10|01}\geq 0\\
c_{1,1}=\mathcal{P}_{1|0}\mathcal{P}_{1|1}-\mathcal{P}_{11|01}\geq 0\\
c_{1,2}=\mathcal{P}_{1|0}-c_{1,0}-c_{1,1}=\mathcal{P}_{1|0}-\mathcal{P}_{1|0}\mathcal{P}_{0|1}+\mathcal{P}_{10|01}-\mathcal{P}_{1|0}\mathcal{P}_{1|1}+\mathcal{P}_{11|01}\geq 0\\
c_{2,0}=\mathcal{P}_{0|1}-c_{0,0}-c_{1,0}=\mathcal{P}_{0|1}-\mathcal{P}_{0|0}\mathcal{P}_{0|1}+\mathcal{P}_{00|01}-\mathcal{P}_{1|0}\mathcal{P}_{0|1}+\mathcal{P}_{10|01}\geq 0\\
c_{2,1}=\mathcal{P}_{1|1}-c_{0,1}-c_{1,1}=\mathcal{P}_{1|1}-\mathcal{P}_{0|0}\mathcal{P}_{1|1}+\mathcal{P}_{01|01}-\mathcal{P}_{1|0}\mathcal{P}_{1|1}+\mathcal{P}_{11|01}\geq 0\\
c_{2,2}=n-\sum\limits_{\forall(a,b)\setminus (2,2)}c_{a,b}=\\
\qquad =n-\mathcal{P}_{0|0}-\mathcal{P}_{1|0}-\mathcal{P}_{0|1}-\mathcal{P}_{1|1}+\mathcal{P}_{0|0}(\mathcal{P}_{0|1}+\mathcal{P}_{1|1})-\mathcal{P}_{00|01}-\mathcal{P}_{01|01}+\mathcal{P}_{1|0}(\mathcal{P}_{0|1}+\mathcal{P}_{1|1})-\mathcal{P}_{10|01}-\mathcal{P}_{11|01}\geq 0 
\end{array}\right.
\end{align}
\end{widetext}
%
%\phantom{.}
%\clearpage
%\newpage

Note that \cref{eq:gEqs} is a set of nine inequality constraints $\{g_i(\vec{\mathcal{P}}_{\boldsymbol{a}|\boldsymbol{x}}) \geq 0\}_{i=1}^9$ and \cref{eq:fEqs} is a set of six equality constraints $\{f_i(\vec{\mathcal{P}}_{\boldsymbol{a}|\boldsymbol{x}}) = 0\}_{i=1}^6$, that altogether define a semialgebraic set $\mathcal{V}$ approximating $\mathbb{P}^S$ from the outside.
This is a valid relaxation because, when $c_{a,a'}$ are all evaluated at an actual integer partition of $n$, the ${\mathcal P}$ polytope coordinates correspond to a projected vertex of ${\mathbb P}$.

Our task is now to certify that the observed statistics lies outside the convex hull of $\mathcal{V}$, which would directly imply that it is also outside of the polytope $P^S$ and thus nonlocal. Deciding membership in the convex hull of a (semi-)algebraic set is a subject of intensive research, but in its full generality it is an NP-hard problem \cite{Lasserre2009}. There exist, however, relaxations to this problem based on hierarchies of SdP \cite{Gouveia2010,Gouveia2012,FadelPRL2017}, which can be efficiently solved using numerical algorithms.

To show how this relaxation is formulated, let us consider the first level of the SdP hierarchy. This consists in taking as basis set $\bm{u}$ the union of $\{1\}$ with the list Eq.~\eqref{eq:Coord14dim}, to then construct ten $15\times 15$-dimensional moment matrices $\Gamma_i = g_i\, \bm{u}\, \bm{u}^T$ for $i=0...9$, with $g_0=1$ and $g_{i>0}$ the expressions in \cref{eq:gEqs}. Combined together, these give the $150\times 150$ block-diagonal matrix $\tilde{\Gamma}$ on which we perform the substitutions \cref{eq:fEqs} in order to impose constraints between the entries. Finally, we linearize $\tilde{\Gamma}$ as $\tilde{\Gamma}(\bm{y})=\sum_i y_i \hat{\Gamma}_i$, where $\hat{\Gamma}_i$ are constant real matrices and $\bm{y}=\{y_i\}_i$ is a list of 616 variables.

The problem of deciding whether the observed statistics $\bm{p}=\{p(\bm{a}|\bm{x})\}_{\bm{a},\bm{x}}$ lies outside $\mathbb{P}^S$ can now be expressed as the SdP \cite{FadelPRL2017}
\begin{equation}
\label{eq:SDPtheta}
\begin{array}{rrcl}
\text{maximize }_{\mathbf{y}, \lambda}& \lambda &&\\
\text{subject to } & \tilde{\Gamma}(\bm{y}) & \succeq & 0 \\
& y_0 & = & 1 \\
& y_j & = & \lambda \, p_j \quad\forall\, j\in\{1,\cdots,\dim(\bm{p})\} \;,
\end{array}
\end{equation}
where some of the variables $y_j$ are set to be proportional to the coordinates of $\bm{p}$, while the rest of $\bm{y}$ components are left as free variables that can be varied to make $\tilde{\Gamma}$ positive semidefinite.

If SdP (\ref{eq:SDPtheta}) returns a value $\lambda \geq 1$, the point $\bm{p}$ is inside the outer approximation of $\mathbb{P}^S$. This does not mean that is it nonlocal, as it might be in the gap between $\mathbb{P}^S$ and its outer approximation. Therefore, the answer is inconclusive and one might need to acess higher orders in the hyerarchy of approximations.

Conversely, if SdP (\ref{eq:SDPtheta}) returns a value $\lambda<1$, the point $\bm{p}$ must be outside the outer approximation of $\mathbb{P}^S$, and it is thus nonlocal. Importantly, the dual formulation of the feasibility version of SdP (\ref{eq:SDPtheta}) results in the dual variables
$\alpha$ associated to $y_1...y_i$, and $\beta_c$ associated to $y_0$, defining a Bell inequality (\ref{eq:gralIneq}) that provides a certificate for the nonlocality of $\bm{p}$.

SdP (\ref{eq:SDPtheta}) can also be used to provide a visualization of the outer approximation to $\mathbb{P}^S$. For small $N$ this can also be compared with a visualization of $\mathbb{P}^S$, thus providing a comparison of the two sets to benchmark the accuracy of our relaxations. To do this, we proceed by choosing two-dimensional (for visualization purposes) planes crossing the local polytope in some arbitrary direction. We then compute the cross section of the local polytope on the plane, as well as the cross section of its outer approximation, and compare the two. Concretely, this is done through the following steps

\begin{enumerate}
\item We select two (random) orthonormal directions $\bm{v}_1, \bm{v}_2$ in the 14-dimensional space (cf. \cref{eq:Coord14dim}) that will define the plane in which the local polytope will be sliced.
\item We set as the origin the point $\bm{v}_{\text{mix}}$ inside the local polytope, which corresponds to the probability distribution of maximal entropy with $p(a_i|x_i)=N/3$ and $p(a_i a_j|x_i x_j)=N(N-1)/9$ for all $i,j$.
\item We parametrize a direction on the plane as $\bm{\tilde{v}}=\cos(\theta)\bm{v}_1+\sin(\theta)\bm{v}_2$.
\item For a given $N$, we write a linear program that checks what is the largest $\lambda$ such that $\lambda \bm{\tilde{v}}$ can be written as a linear combination of the vertices of ${\mathbbm{P}}^S$.
%For small $N$, we find the boundary of $\mathbb{P}^S$ along direction $\bm{p}$ with length $\lambda$ given by \cref{eq:SDPtheta}.
% Look for feasibility with the following Linear Programming problem in order to find the polytope boundary: \mf{IT IS NOT THIS, WE DO SOMETHING SIMILAR TO THE SDP WITH LAMBDA}
% \begin{equation}
% \begin{array}{rccl}
%    \text{minimize}& \bm{0}^T \bm{x} && \\
%     \text{subject to}& [A,(0,\bm{p})^T]\bm{x} &=& \bm{v}_{\text{mix}} \\
%     & \bm{x} &\geq& 0 \\
%     & \bm{x} &\leq& 1, 
%     \end{array}
% \end{equation}
% where $\bm{x}$ is the LP free variable, $\bm{0}$ denotes the zero column vector and $[A,(0,\bm{p})^T]$ denotes a matrix where its submatrix $A$ is constructed from the column vectors corresponding to all the possible combinations in \Cref{table:LDS}.
\item We find the boundary of the approximation to the local polytope along direction $\bm{\tilde{v}}$ as the solution to SdP (\ref{eq:SDPtheta}) with now $y_j = \lambda \bm{\tilde{v}}_j + \left(\bm{v}_{\text{mix}} \right)_j, \quad\forall\, j\in\{1,\cdots,\dim(\bm{\tilde{v}})\}$. 
\item We repeat steps 3, 4 and 5 for several values of $\theta\in\{0,\ldots,2\pi\}$ until a full sweep has been completed.
\end{enumerate}

\begin{figure*}[htpb]
\centering
\includegraphics[width=0.99\textwidth]{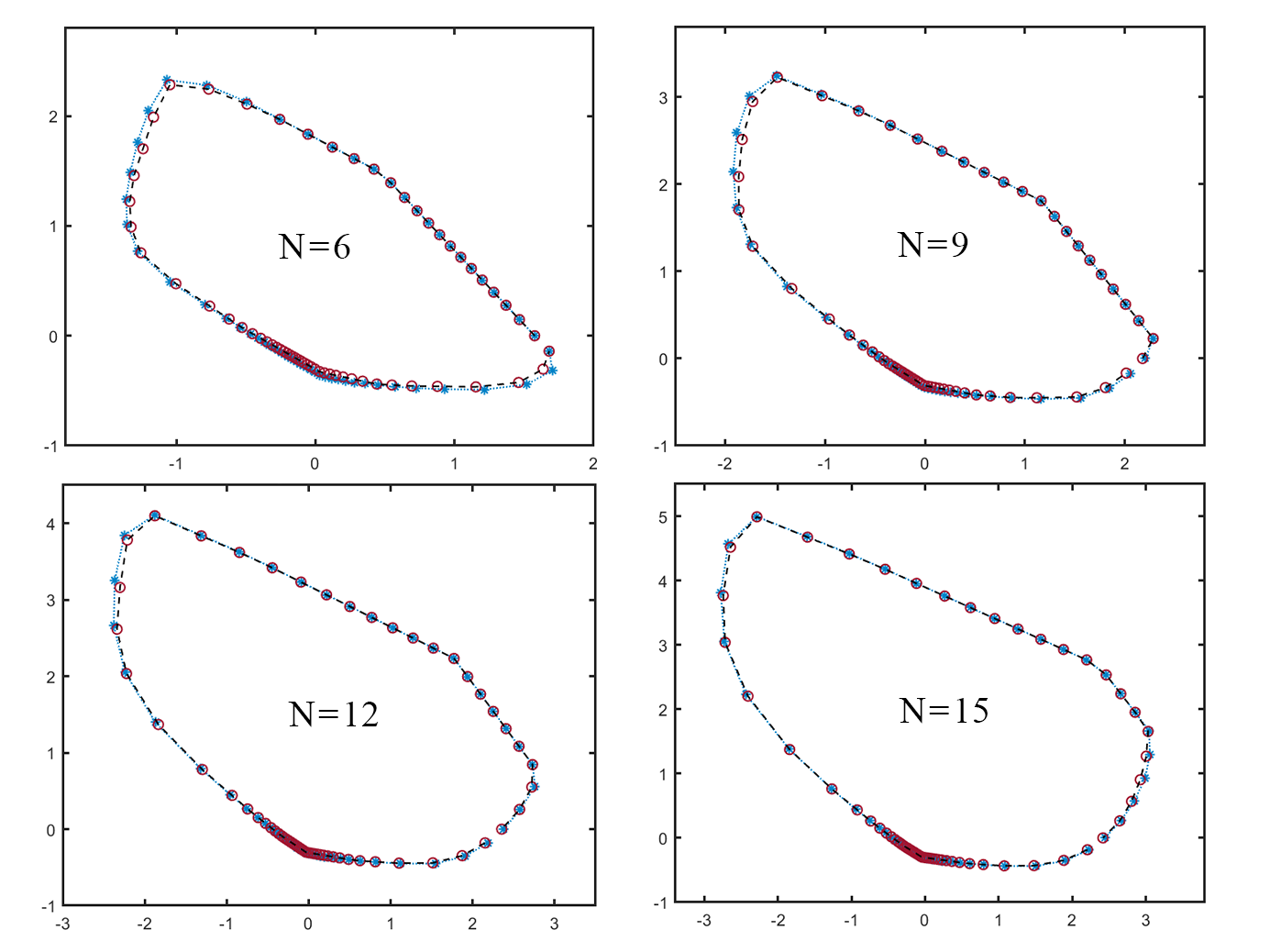}
\caption{Slices of the local polytope $\mathbb{P}^S$ (red circles and black dashed  line), compared to a slice of its outer approximation provided by SdP~(\ref{eq:SDPtheta}) (blue asterisks and blue dotted line), for different number of parties $N$. These slices are taken on a plane defined by two orthonormal directions $\bm{v}_1, \bm{v}_2$ randomly chosen in the 14-dimensional space spanned by \cref{eq:Coord14dim}. One observes that the outer approximation becomes relatively tighter as $N$ increases \cite{FadelPRL2017}. 
%\al{The directions for this figure are $\bm{v}_1=(-0.2045,0.1541,0.0457,-0.3860,-0.1409,0.3596, 0.1613,-0.0720,-0.2788,-0.0621,-0.3666,-0.0488, 0.2705,-0.0856)$ and $\bm{v}_2=(-0.4597,-0.1009,-0.2348,-0.2459,-0.1323,-0.3129,0.4401,-0.4856,0.0008,-0.0247,0.1191,0.2800,-0.0900,0.4317).$}
} 
\label{fig:potato}
\end{figure*}

We show in \Cref{fig:potato} examples of slices taken in different planes and for different $N$. We see that, already for relatively small $N$, our SdP method tightly approximates the local polytope $\mathbb{P}^S$. Crucially, this approximation is expected to improve even further as $N$ increases \cite{FadelPRL2017}.

\section{Conclusions}
We discussed the problem of deriving new multipartite Bell inequalities in the $(N,2,3)$ scenario, namely when each of the $N$ parties chooses to perform one out of two possible three-outcome measurements. 
As the complexity of fully characterizing the set of classical correlations becomes quickly intractable already for $N>2$, we consider a simplification to the problem: We focus on deriving Bell inequalities involving permutationally invariant one- and two-body observables. 
This is geometrically understood as characterizing only a projection of the full set of classical correlations onto a low-dimensional subspace. Importantly, imposing permutational invariance results in the dimension of this subspace being independent of the number of parties $N$.

We propose two complementary approaches for deriving  3-outcome permutationally invariant Bell inequalities (3PIBIs). The first consists in deriving 3PIBIs for small $N\leq 17$, to then infer their possible generalization for arbitrary large $N$. The second consists in approximating the projected set of classical correlations from the outside with a hierarchy of SdPs \cite{FadelPRL2017}, which provides a certificate for non-membership of a point in the set taking the form of a valid 3PIBI. 

The tools we propose have application in deriving new Bell inequalities for the detection of nonlocality in paradigmatic three-level Hamiltonians \cite{MeredithPRA1988,GnutzmannJPhysA1999}, as well as for the construction of new experimentally-friendly Bell correlation witnesses suited to ensembles of spin-1 particles \cite{HamleyNat12,OurFriends}. These could enable the experimental detection of high-dimensional correlations in multipartite systems, that are genuinely distinct from their two-level counterparts \cite{UsInPrep2}.

\vspace{2mm}
\noindent\textbf{Code and data availability}.--  The code and data to construct the plots are available at the following link, \url{https://github.com/Albert-Aloy/3PIBIs}.

%\vspace{2mm}
\noindent\textbf{Acknowledgments}.-- 
AA acknowledges support from the Austrian Science Fund (FWF) (projects P 33730-N and 10.55776/PAT2839723) and by the ESQ Discovery programme (Erwin Schr{\"o}dinger Center for Quantum Science \& Technology), hosted by the Austrian Academy of Sciences ({\"O}AW).

GM acknowledges support from: Europea Research Council AdG NOQIA; MCIN/AEI (PGC2018-0910.13039/501100011033, CEX2019-000910-S/10.13039/501100011033, Plan National FIDEUA PID2019-106901GB-I00, Plan National STAMEENA PID2022-139099NB, I00, project funded by MCIN/AEI/10.13039/501100011033 and by the “European Union NextGenerationEU/PRTR" (PRTR-C17.I1), FPI); QUANTERA MAQS PCI2019-111828-2); QUANTERA DYNAMITE PCI2022-132919, QuantERA II Programme co-funded by European Union’s Horizon 2020 program under Grant Agreement No 101017733); Ministry for Digital Transformation and of Civil Service of the Spanish Government through the QUANTUM ENIA project call - Quantum Spain project, and by the European Union through the Recovery, Transformation and Resilience Plan - NextGenerationEU within the framework of the Digital Spain 2026 Agenda; Fundació Cellex; Fundació Mir-Puig; Generalitat de Catalunya (European Social Fund FEDER and CERCA program, AGAUR Grant No. 2021 SGR 01452, QuantumCAT \ U16-011424, co-funded by ERDF Operational Program of Catalonia 2014-2020); Barcelona Supercomputing Center MareNostrum (FI-2023-3-0024); Funded by the European Union. Views and opinions expressed are however those of the author(s) only and do not necessarily reflect those of the European Union, European Commission, European Climate, Infrastructure and Environment Executive Agency (CINEA), or any other granting authority. Neither the European Union nor any granting authority can be held responsible for them (HORIZON-CL4-2022-QUANTUM-02-SGA PASQuanS2.1, 101113690, EU Horizon 2020 FET-OPEN OPTOlogic, Grant No 899794), EU Horizon Europe Program (This project has received funding from the European Union’s Horizon Europe research and innovation program under grant agreement No 101080086 NeQSTGrant Agreement 101080086 — NeQST); ICFO Internal “QuantumGaudi” project; European Union’s Horizon 2020 program under the Marie Sklodowska-Curie grant agreement No 847648; “La Caixa” Junior Leaders fellowships, La Caixa” Foundation (ID 100010434): CF/BQ/PR23/11980043.

JT has received support from the European Union’s Horizon Europe program through the ERC StG FINE-TEA-SQUAD (Grant No. 101040729). JT also acknowledges support from the Quantum Delta NL program. This publication is part of the ‘Quantum Inspire – the Dutch Quantum Computer in the Cloud’ project (with project number [NWA.1292.19.194]) of the NWA research program ‘Research on Routes by Consortia (ORC)’, which is funded by the Netherlands Organization for Scientific Research (NWO).

MF was supported by the Swiss National Science Foundation Ambizione Grant No. 208886, and The Branco Weiss Fellowship -- Society in Science, administered by the ETH Z\"{u}rich.

\bibliographystyle{apsrev4-1} % Tell bibtex which bibliography style to use
\bibliography{BellDim.bib}

\begin{thebibliography}{43}%
\makeatletter
\providecommand \@ifxundefined [1]{%
 \@ifx{#1\undefined}
}%
\providecommand \@ifnum [1]{%
 \ifnum #1\expandafter \@firstoftwo
 \else \expandafter \@secondoftwo
 \fi
}%
\providecommand \@ifx [1]{%
 \ifx #1\expandafter \@firstoftwo
 \else \expandafter \@secondoftwo
 \fi
}%
\providecommand \natexlab [1]{#1}%
\providecommand \enquote  [1]{``#1''}%
\providecommand \bibnamefont  [1]{#1}%
\providecommand \bibfnamefont [1]{#1}%
\providecommand \citenamefont [1]{#1}%
\providecommand \href@noop [0]{\@secondoftwo}%
\providecommand \href [0]{\begingroup \@sanitize@url \@href}%
\providecommand \@href[1]{\@@startlink{#1}\@@href}%
\providecommand \@@href[1]{\endgroup#1\@@endlink}%
\providecommand \@sanitize@url [0]{\catcode `\\12\catcode `\$12\catcode `\&12\catcode `\#12\catcode `\^12\catcode `\_12\catcode `\%12\relax}%
\providecommand \@@startlink[1]{}%
\providecommand \@@endlink[0]{}%
\providecommand \url  [0]{\begingroup\@sanitize@url \@url }%
\providecommand \@url [1]{\endgroup\@href {#1}{\urlprefix }}%
\providecommand \urlprefix  [0]{URL }%
\providecommand \Eprint [0]{\href }%
\providecommand \doibase [0]{http://dx.doi.org/}%
\providecommand \selectlanguage [0]{\@gobble}%
\providecommand \bibinfo  [0]{\@secondoftwo}%
\providecommand \bibfield  [0]{\@secondoftwo}%
\providecommand \translation [1]{[#1]}%
\providecommand \BibitemOpen [0]{}%
\providecommand \bibitemStop [0]{}%
\providecommand \bibitemNoStop [0]{.\EOS\space}%
\providecommand \EOS [0]{\spacefactor3000\relax}%
\providecommand \BibitemShut  [1]{\csname bibitem#1\endcsname}%
\let\auto@bib@innerbib\@empty
%</preamble>
\bibitem [{\citenamefont {Brunner}\ \emph {et~al.}(2014)\citenamefont {Brunner}, \citenamefont {Cavalcanti}, \citenamefont {Pironio}, \citenamefont {Scarani},\ and\ \citenamefont {Wehner}}]{BrunnerRMP2014}%
  \BibitemOpen
  \bibfield  {author} {\bibinfo {author} {\bibfnamefont {N.}~\bibnamefont {Brunner}}, \bibinfo {author} {\bibfnamefont {D.}~\bibnamefont {Cavalcanti}}, \bibinfo {author} {\bibfnamefont {S.}~\bibnamefont {Pironio}}, \bibinfo {author} {\bibfnamefont {V.}~\bibnamefont {Scarani}}, \ and\ \bibinfo {author} {\bibfnamefont {S.}~\bibnamefont {Wehner}},\ }\href {\doibase 10.1103/RevModPhys.86.419} {\bibfield  {journal} {\bibinfo  {journal} {Rev. Mod. Phys.}\ }\textbf {\bibinfo {volume} {86}},\ \bibinfo {pages} {419} (\bibinfo {year} {2014})}\BibitemShut {NoStop}%
\bibitem [{\citenamefont {Frérot}\ \emph {et~al.}(2023)\citenamefont {Frérot}, \citenamefont {Fadel},\ and\ \citenamefont {Lewenstein}}]{FrerotROPP2023}%
  \BibitemOpen
  \bibfield  {author} {\bibinfo {author} {\bibfnamefont {I.}~\bibnamefont {Frérot}}, \bibinfo {author} {\bibfnamefont {M.}~\bibnamefont {Fadel}}, \ and\ \bibinfo {author} {\bibfnamefont {M.}~\bibnamefont {Lewenstein}},\ }\href {\doibase 10.1088/1361-6633/acf8d7} {\bibfield  {journal} {\bibinfo  {journal} {Reports on Progress in Physics}\ }\textbf {\bibinfo {volume} {86}},\ \bibinfo {pages} {114001} (\bibinfo {year} {2023})}\BibitemShut {NoStop}%
\bibitem [{\citenamefont {Babai}\ \emph {et~al.}(1991)\citenamefont {Babai}, \citenamefont {Fortnow},\ and\ \citenamefont {Lund}}]{BabaiCompCompl1991}%
  \BibitemOpen
  \bibfield  {author} {\bibinfo {author} {\bibfnamefont {L.}~\bibnamefont {Babai}}, \bibinfo {author} {\bibfnamefont {L.}~\bibnamefont {Fortnow}}, \ and\ \bibinfo {author} {\bibfnamefont {C.}~\bibnamefont {Lund}},\ }\href {\doibase 10.1007/bf01200056} {\bibfield  {journal} {\bibinfo  {journal} {Computational Complexity}\ }\textbf {\bibinfo {volume} {1}},\ \bibinfo {pages} {3} (\bibinfo {year} {1991})}\BibitemShut {NoStop}%
\bibitem [{\citenamefont {Tura}\ \emph {et~al.}(2014{\natexlab{a}})\citenamefont {Tura}, \citenamefont {Sainz}, \citenamefont {Vértesi}, \citenamefont {Acín}, \citenamefont {Lewenstein},\ and\ \citenamefont {Augusiak}}]{Tura_2014}%
  \BibitemOpen
  \bibfield  {author} {\bibinfo {author} {\bibfnamefont {J.}~\bibnamefont {Tura}}, \bibinfo {author} {\bibfnamefont {A.~B.}\ \bibnamefont {Sainz}}, \bibinfo {author} {\bibfnamefont {T.}~\bibnamefont {Vértesi}}, \bibinfo {author} {\bibfnamefont {A.}~\bibnamefont {Acín}}, \bibinfo {author} {\bibfnamefont {M.}~\bibnamefont {Lewenstein}}, \ and\ \bibinfo {author} {\bibfnamefont {R.}~\bibnamefont {Augusiak}},\ }\href {\doibase 10.1088/1751-8113/47/42/424024} {\bibfield  {journal} {\bibinfo  {journal} {Journal of Physics A: Mathematical and Theoretical}\ }\textbf {\bibinfo {volume} {47}},\ \bibinfo {pages} {424024} (\bibinfo {year} {2014}{\natexlab{a}})}\BibitemShut {NoStop}%
\bibitem [{\citenamefont {Wang}\ \emph {et~al.}(2017)\citenamefont {Wang}, \citenamefont {Singh},\ and\ \citenamefont {Navascu\'es}}]{WangPRL2017}%
  \BibitemOpen
  \bibfield  {author} {\bibinfo {author} {\bibfnamefont {Z.}~\bibnamefont {Wang}}, \bibinfo {author} {\bibfnamefont {S.}~\bibnamefont {Singh}}, \ and\ \bibinfo {author} {\bibfnamefont {M.}~\bibnamefont {Navascu\'es}},\ }\href {\doibase 10.1103/PhysRevLett.118.230401} {\bibfield  {journal} {\bibinfo  {journal} {Phys. Rev. Lett.}\ }\textbf {\bibinfo {volume} {118}},\ \bibinfo {pages} {230401} (\bibinfo {year} {2017})}\BibitemShut {NoStop}%
\bibitem [{\citenamefont {Tura}\ \emph {et~al.}(2014{\natexlab{b}})\citenamefont {Tura}, \citenamefont {Augusiak}, \citenamefont {Sainz}, \citenamefont {V{\'e}rtesi}, \citenamefont {Lewenstein},\ and\ \citenamefont {Ac{\'\i}n}}]{SciencePaper}%
  \BibitemOpen
  \bibfield  {author} {\bibinfo {author} {\bibfnamefont {J.}~\bibnamefont {Tura}}, \bibinfo {author} {\bibfnamefont {R.}~\bibnamefont {Augusiak}}, \bibinfo {author} {\bibfnamefont {A.~B.}\ \bibnamefont {Sainz}}, \bibinfo {author} {\bibfnamefont {T.}~\bibnamefont {V{\'e}rtesi}}, \bibinfo {author} {\bibfnamefont {M.}~\bibnamefont {Lewenstein}}, \ and\ \bibinfo {author} {\bibfnamefont {A.}~\bibnamefont {Ac{\'\i}n}},\ }\href {\doibase 10.1126/science.1247715} {\bibfield  {journal} {\bibinfo  {journal} {Science}\ }\textbf {\bibinfo {volume} {344}},\ \bibinfo {pages} {1256} (\bibinfo {year} {2014}{\natexlab{b}})}\BibitemShut {NoStop}%
\bibitem [{\citenamefont {Wagner}\ \emph {et~al.}(2017)\citenamefont {Wagner}, \citenamefont {Schmied}, \citenamefont {Fadel}, \citenamefont {Treutlein}, \citenamefont {Sangouard},\ and\ \citenamefont {Bancal}}]{WagnerPRL2017}%
  \BibitemOpen
  \bibfield  {author} {\bibinfo {author} {\bibfnamefont {S.}~\bibnamefont {Wagner}}, \bibinfo {author} {\bibfnamefont {R.}~\bibnamefont {Schmied}}, \bibinfo {author} {\bibfnamefont {M.}~\bibnamefont {Fadel}}, \bibinfo {author} {\bibfnamefont {P.}~\bibnamefont {Treutlein}}, \bibinfo {author} {\bibfnamefont {N.}~\bibnamefont {Sangouard}}, \ and\ \bibinfo {author} {\bibfnamefont {J.-D.}\ \bibnamefont {Bancal}},\ }\href {\doibase 10.1103/physrevlett.119.170403} {\bibfield  {journal} {\bibinfo  {journal} {Physical Review Letters}\ }\textbf {\bibinfo {volume} {119}},\ \bibinfo {pages} {170403} (\bibinfo {year} {2017})}\BibitemShut {NoStop}%
\bibitem [{\citenamefont {Baccari}\ \emph {et~al.}(2019)\citenamefont {Baccari}, \citenamefont {Tura}, \citenamefont {Fadel}, \citenamefont {Aloy}, \citenamefont {Bancal}, \citenamefont {Sangouard}, \citenamefont {Lewenstein}, \citenamefont {Ac{\'{\i}}n},\ and\ \citenamefont {Augusiak}}]{BaccariPRA2019}%
  \BibitemOpen
  \bibfield  {author} {\bibinfo {author} {\bibfnamefont {F.}~\bibnamefont {Baccari}}, \bibinfo {author} {\bibfnamefont {J.}~\bibnamefont {Tura}}, \bibinfo {author} {\bibfnamefont {M.}~\bibnamefont {Fadel}}, \bibinfo {author} {\bibfnamefont {A.}~\bibnamefont {Aloy}}, \bibinfo {author} {\bibfnamefont {J.-D.}\ \bibnamefont {Bancal}}, \bibinfo {author} {\bibfnamefont {N.}~\bibnamefont {Sangouard}}, \bibinfo {author} {\bibfnamefont {M.}~\bibnamefont {Lewenstein}}, \bibinfo {author} {\bibfnamefont {A.}~\bibnamefont {Ac{\'{\i}}n}}, \ and\ \bibinfo {author} {\bibfnamefont {R.}~\bibnamefont {Augusiak}},\ }\href {10.1103/physreva.100.022121} {\bibfield  {journal} {\bibinfo  {journal} {Physical Review A}\ }\textbf {\bibinfo {volume} {100}} (\bibinfo {year} {2019})}\BibitemShut {NoStop}%
\bibitem [{\citenamefont {Fadel}\ and\ \citenamefont {Hern\'andez-Cuenca}(2022)}]{FadelPRD22}%
  \BibitemOpen
  \bibfield  {author} {\bibinfo {author} {\bibfnamefont {M.}~\bibnamefont {Fadel}}\ and\ \bibinfo {author} {\bibfnamefont {S.}~\bibnamefont {Hern\'andez-Cuenca}},\ }\href {\doibase 10.1103/PhysRevD.105.086008} {\bibfield  {journal} {\bibinfo  {journal} {Phys. Rev. D}\ }\textbf {\bibinfo {volume} {105}},\ \bibinfo {pages} {086008} (\bibinfo {year} {2022})}\BibitemShut {NoStop}%
\bibitem [{\citenamefont {Guo}\ \emph {et~al.}(2023)\citenamefont {Guo}, \citenamefont {Tura}, \citenamefont {He},\ and\ \citenamefont {Fadel}}]{GuoPRL23}%
  \BibitemOpen
  \bibfield  {author} {\bibinfo {author} {\bibfnamefont {J.}~\bibnamefont {Guo}}, \bibinfo {author} {\bibfnamefont {J.}~\bibnamefont {Tura}}, \bibinfo {author} {\bibfnamefont {Q.}~\bibnamefont {He}}, \ and\ \bibinfo {author} {\bibfnamefont {M.}~\bibnamefont {Fadel}},\ }\href {\doibase 10.1103/PhysRevLett.131.070201} {\bibfield  {journal} {\bibinfo  {journal} {Phys. Rev. Lett.}\ }\textbf {\bibinfo {volume} {131}},\ \bibinfo {pages} {070201} (\bibinfo {year} {2023})}\BibitemShut {NoStop}%
\bibitem [{\citenamefont {Fadel}\ and\ \citenamefont {Tura}(2018)}]{FadelQuantum2018}%
  \BibitemOpen
  \bibfield  {author} {\bibinfo {author} {\bibfnamefont {M.}~\bibnamefont {Fadel}}\ and\ \bibinfo {author} {\bibfnamefont {J.}~\bibnamefont {Tura}},\ }\href {\doibase 10.22331/q-2018-11-19-107} {\bibfield  {journal} {\bibinfo  {journal} {{Quantum}}\ }\textbf {\bibinfo {volume} {2}},\ \bibinfo {pages} {107} (\bibinfo {year} {2018})}\BibitemShut {NoStop}%
\bibitem [{\citenamefont {Piga}\ \emph {et~al.}(2019)\citenamefont {Piga}, \citenamefont {Aloy}, \citenamefont {Lewenstein},\ and\ \citenamefont {Fr{\'{e}}rot}}]{PigaPRL2019}%
  \BibitemOpen
  \bibfield  {author} {\bibinfo {author} {\bibfnamefont {A.}~\bibnamefont {Piga}}, \bibinfo {author} {\bibfnamefont {A.}~\bibnamefont {Aloy}}, \bibinfo {author} {\bibfnamefont {M.}~\bibnamefont {Lewenstein}}, \ and\ \bibinfo {author} {\bibfnamefont {I.}~\bibnamefont {Fr{\'{e}}rot}},\ }\href {10.1103/physrevlett.123.170604} {\bibfield  {journal} {\bibinfo  {journal} {Physical Review Letters}\ }\textbf {\bibinfo {volume} {123}} (\bibinfo {year} {2019})}\BibitemShut {NoStop}%
\bibitem [{\citenamefont {Fröwis}\ \emph {et~al.}(2019)\citenamefont {Fröwis}, \citenamefont {Fadel}, \citenamefont {Treutlein}, \citenamefont {Gisin},\ and\ \citenamefont {Brunner}}]{FroewisPRAR2019}%
  \BibitemOpen
  \bibfield  {author} {\bibinfo {author} {\bibfnamefont {F.}~\bibnamefont {Fröwis}}, \bibinfo {author} {\bibfnamefont {M.}~\bibnamefont {Fadel}}, \bibinfo {author} {\bibfnamefont {P.}~\bibnamefont {Treutlein}}, \bibinfo {author} {\bibfnamefont {N.}~\bibnamefont {Gisin}}, \ and\ \bibinfo {author} {\bibfnamefont {N.}~\bibnamefont {Brunner}},\ }\href {\doibase 10.1103/physreva.99.040101} {\bibfield  {journal} {\bibinfo  {journal} {Physical Review A}\ }\textbf {\bibinfo {volume} {99}} (\bibinfo {year} {2019}),\ 10.1103/physreva.99.040101}\BibitemShut {NoStop}%
\bibitem [{\citenamefont {Schmied}\ \emph {et~al.}(2016)\citenamefont {Schmied}, \citenamefont {Bancal}, \citenamefont {Allard}, \citenamefont {Fadel}, \citenamefont {Scarani}, \citenamefont {Treutlein},\ and\ \citenamefont {Sangouard}}]{SchmiedScience2016}%
  \BibitemOpen
  \bibfield  {author} {\bibinfo {author} {\bibfnamefont {R.}~\bibnamefont {Schmied}}, \bibinfo {author} {\bibfnamefont {J.-D.}\ \bibnamefont {Bancal}}, \bibinfo {author} {\bibfnamefont {B.}~\bibnamefont {Allard}}, \bibinfo {author} {\bibfnamefont {M.}~\bibnamefont {Fadel}}, \bibinfo {author} {\bibfnamefont {V.}~\bibnamefont {Scarani}}, \bibinfo {author} {\bibfnamefont {P.}~\bibnamefont {Treutlein}}, \ and\ \bibinfo {author} {\bibfnamefont {N.}~\bibnamefont {Sangouard}},\ }\href {\doibase 10.1126/science.aad8665} {\bibfield  {journal} {\bibinfo  {journal} {Science}\ }\textbf {\bibinfo {volume} {352}},\ \bibinfo {pages} {441} (\bibinfo {year} {2016})}\BibitemShut {NoStop}%
\bibitem [{\citenamefont {Engelsen}\ \emph {et~al.}(2017)\citenamefont {Engelsen}, \citenamefont {Krishnakumar}, \citenamefont {Hosten},\ and\ \citenamefont {Kasevich}}]{EngelsenPRL2017}%
  \BibitemOpen
  \bibfield  {author} {\bibinfo {author} {\bibfnamefont {N.~J.}\ \bibnamefont {Engelsen}}, \bibinfo {author} {\bibfnamefont {R.}~\bibnamefont {Krishnakumar}}, \bibinfo {author} {\bibfnamefont {O.}~\bibnamefont {Hosten}}, \ and\ \bibinfo {author} {\bibfnamefont {M.~A.}\ \bibnamefont {Kasevich}},\ }\href {\doibase 10.1103/PhysRevLett.118.140401} {\bibfield  {journal} {\bibinfo  {journal} {Phys. Rev. Lett.}\ }\textbf {\bibinfo {volume} {118}},\ \bibinfo {pages} {140401} (\bibinfo {year} {2017})}\BibitemShut {NoStop}%
\bibitem [{\citenamefont {Alsina}\ \emph {et~al.}(2016)\citenamefont {Alsina}, \citenamefont {Cervera}, \citenamefont {Goyeneche}, \citenamefont {Latorre},\ and\ \citenamefont {\ifmmode~\dot{Z}\else \.{Z}\fi{}yczkowski}}]{AlsinaPRA16}%
  \BibitemOpen
  \bibfield  {author} {\bibinfo {author} {\bibfnamefont {D.}~\bibnamefont {Alsina}}, \bibinfo {author} {\bibfnamefont {A.}~\bibnamefont {Cervera}}, \bibinfo {author} {\bibfnamefont {D.}~\bibnamefont {Goyeneche}}, \bibinfo {author} {\bibfnamefont {J.~I.}\ \bibnamefont {Latorre}}, \ and\ \bibinfo {author} {\bibfnamefont {K.}~\bibnamefont {\ifmmode~\dot{Z}\else \.{Z}\fi{}yczkowski}},\ }\href {\doibase 10.1103/PhysRevA.94.032102} {\bibfield  {journal} {\bibinfo  {journal} {Phys. Rev. A}\ }\textbf {\bibinfo {volume} {94}},\ \bibinfo {pages} {032102} (\bibinfo {year} {2016})}\BibitemShut {NoStop}%
\bibitem [{\citenamefont {M\"uller-Rigat}\ \emph {et~al.}(2021)\citenamefont {M\"uller-Rigat}, \citenamefont {Aloy}, \citenamefont {Lewenstein},\ and\ \citenamefont {Frèrot}}]{GuillemPRXQuantum}%
  \BibitemOpen
  \bibfield  {author} {\bibinfo {author} {\bibfnamefont {G.}~\bibnamefont {M\"uller-Rigat}}, \bibinfo {author} {\bibfnamefont {A.}~\bibnamefont {Aloy}}, \bibinfo {author} {\bibfnamefont {M.}~\bibnamefont {Lewenstein}}, \ and\ \bibinfo {author} {\bibfnamefont {I.}~\bibnamefont {Frèrot}},\ }\href {\doibase 10.1103/PRXQuantum.2.030329} {\bibfield  {journal} {\bibinfo  {journal} {PRX Quantum}\ }\textbf {\bibinfo {volume} {2}},\ \bibinfo {pages} {030329} (\bibinfo {year} {2021})}\BibitemShut {NoStop}%
\bibitem [{\citenamefont {Lipkin}\ \emph {et~al.}(1965)\citenamefont {Lipkin}, \citenamefont {Meshkov},\ and\ \citenamefont {Glick}}]{LipkinNucPhys1965}%
  \BibitemOpen
  \bibfield  {author} {\bibinfo {author} {\bibfnamefont {H.}~\bibnamefont {Lipkin}}, \bibinfo {author} {\bibfnamefont {N.}~\bibnamefont {Meshkov}}, \ and\ \bibinfo {author} {\bibfnamefont {A.}~\bibnamefont {Glick}},\ }\href {\doibase 10.1016/0029-5582(65)90862-x} {\bibfield  {journal} {\bibinfo  {journal} {Nuclear Physics}\ }\textbf {\bibinfo {volume} {62}},\ \bibinfo {pages} {188} (\bibinfo {year} {1965})}\BibitemShut {NoStop}%
\bibitem [{\citenamefont {Meshkov}\ \emph {et~al.}(1965)\citenamefont {Meshkov}, \citenamefont {Glick},\ and\ \citenamefont {Lipkin}}]{MeshkovNucPhys1965}%
  \BibitemOpen
  \bibfield  {author} {\bibinfo {author} {\bibfnamefont {N.}~\bibnamefont {Meshkov}}, \bibinfo {author} {\bibfnamefont {A.}~\bibnamefont {Glick}}, \ and\ \bibinfo {author} {\bibfnamefont {H.}~\bibnamefont {Lipkin}},\ }\href {\doibase 10.1016/0029-5582(65)90863-1} {\bibfield  {journal} {\bibinfo  {journal} {Nuclear Physics}\ }\textbf {\bibinfo {volume} {62}},\ \bibinfo {pages} {199} (\bibinfo {year} {1965})}\BibitemShut {NoStop}%
\bibitem [{\citenamefont {Glick}\ \emph {et~al.}(1965)\citenamefont {Glick}, \citenamefont {Lipkin},\ and\ \citenamefont {Meshkov}}]{GlickNucPhys1965}%
  \BibitemOpen
  \bibfield  {author} {\bibinfo {author} {\bibfnamefont {A.}~\bibnamefont {Glick}}, \bibinfo {author} {\bibfnamefont {H.}~\bibnamefont {Lipkin}}, \ and\ \bibinfo {author} {\bibfnamefont {N.}~\bibnamefont {Meshkov}},\ }\href {\doibase 10.1016/0029-5582(65)90864-3} {\bibfield  {journal} {\bibinfo  {journal} {Nuclear Physics}\ }\textbf {\bibinfo {volume} {62}},\ \bibinfo {pages} {211} (\bibinfo {year} {1965})}\BibitemShut {NoStop}%
\bibitem [{\citenamefont {Meredith}\ \emph {et~al.}(1988)\citenamefont {Meredith}, \citenamefont {Koonin},\ and\ \citenamefont {Zirnbauer}}]{MeredithPRA1988}%
  \BibitemOpen
  \bibfield  {author} {\bibinfo {author} {\bibfnamefont {D.~C.}\ \bibnamefont {Meredith}}, \bibinfo {author} {\bibfnamefont {S.~E.}\ \bibnamefont {Koonin}}, \ and\ \bibinfo {author} {\bibfnamefont {M.~R.}\ \bibnamefont {Zirnbauer}},\ }\href {\doibase 10.1103/physreva.37.3499} {\bibfield  {journal} {\bibinfo  {journal} {Physical Review A}\ }\textbf {\bibinfo {volume} {37}},\ \bibinfo {pages} {3499} (\bibinfo {year} {1988})}\BibitemShut {NoStop}%
\bibitem [{\citenamefont {Law}\ \emph {et~al.}(1998)\citenamefont {Law}, \citenamefont {Pu},\ and\ \citenamefont {Bigelow}}]{LawPRL98}%
  \BibitemOpen
  \bibfield  {author} {\bibinfo {author} {\bibfnamefont {C.~K.}\ \bibnamefont {Law}}, \bibinfo {author} {\bibfnamefont {H.}~\bibnamefont {Pu}}, \ and\ \bibinfo {author} {\bibfnamefont {N.~P.}\ \bibnamefont {Bigelow}},\ }\href {\doibase 10.1103/PhysRevLett.81.5257} {\bibfield  {journal} {\bibinfo  {journal} {Phys. Rev. Lett.}\ }\textbf {\bibinfo {volume} {81}},\ \bibinfo {pages} {5257} (\bibinfo {year} {1998})}\BibitemShut {NoStop}%
\bibitem [{\citenamefont {Haldane}(1983{\natexlab{a}})}]{haldane1983continuum}%
  \BibitemOpen
  \bibfield  {author} {\bibinfo {author} {\bibfnamefont {F.~D.~M.}\ \bibnamefont {Haldane}},\ }\href {\doibase 10.1016/0375-9601(83)90631-X} {\bibfield  {journal} {\bibinfo  {journal} {Physics Letters A}\ }\textbf {\bibinfo {volume} {93}},\ \bibinfo {pages} {464} (\bibinfo {year} {1983}{\natexlab{a}})}\BibitemShut {NoStop}%
\bibitem [{\citenamefont {Haldane}(1983{\natexlab{b}})}]{HaldanePRL83}%
  \BibitemOpen
  \bibfield  {author} {\bibinfo {author} {\bibfnamefont {F.~D.~M.}\ \bibnamefont {Haldane}},\ }\href {\doibase 10.1103/PhysRevLett.50.1153} {\bibfield  {journal} {\bibinfo  {journal} {Phys. Rev. Lett.}\ }\textbf {\bibinfo {volume} {50}},\ \bibinfo {pages} {1153} (\bibinfo {year} {1983}{\natexlab{b}})}\BibitemShut {NoStop}%
\bibitem [{\citenamefont {Affleck}\ \emph {et~al.}(1987)\citenamefont {Affleck}, \citenamefont {Kennedy}, \citenamefont {Lieb},\ and\ \citenamefont {Tasaki}}]{AKLTPRL87}%
  \BibitemOpen
  \bibfield  {author} {\bibinfo {author} {\bibfnamefont {I.}~\bibnamefont {Affleck}}, \bibinfo {author} {\bibfnamefont {T.}~\bibnamefont {Kennedy}}, \bibinfo {author} {\bibfnamefont {E.~H.}\ \bibnamefont {Lieb}}, \ and\ \bibinfo {author} {\bibfnamefont {H.}~\bibnamefont {Tasaki}},\ }\href {\doibase 10.1103/PhysRevLett.59.799} {\bibfield  {journal} {\bibinfo  {journal} {Phys. Rev. Lett.}\ }\textbf {\bibinfo {volume} {59}},\ \bibinfo {pages} {799} (\bibinfo {year} {1987})}\BibitemShut {NoStop}%
\bibitem [{\citenamefont {Gnutzmann}\ \emph {et~al.}(1999)\citenamefont {Gnutzmann}, \citenamefont {Haake},\ and\ \citenamefont {Kus}}]{GnutzmannJPhysA1999}%
  \BibitemOpen
  \bibfield  {author} {\bibinfo {author} {\bibfnamefont {S.}~\bibnamefont {Gnutzmann}}, \bibinfo {author} {\bibfnamefont {F.}~\bibnamefont {Haake}}, \ and\ \bibinfo {author} {\bibfnamefont {M.}~\bibnamefont {Kus}},\ }\href {\doibase 10.1088/0305-4470/33/1/309} {\bibfield  {journal} {\bibinfo  {journal} {Journal of Physics A: Mathematical and General}\ }\textbf {\bibinfo {volume} {33}},\ \bibinfo {pages} {143} (\bibinfo {year} {1999})}\BibitemShut {NoStop}%
\bibitem [{\citenamefont {Hamley}\ \emph {et~al.}(2012)\citenamefont {Hamley}, \citenamefont {Gerving}, \citenamefont {Hoang}, \citenamefont {Bookjans},\ and\ \citenamefont {Chapman}}]{HamleyNat12}%
  \BibitemOpen
  \bibfield  {author} {\bibinfo {author} {\bibfnamefont {C.~D.}\ \bibnamefont {Hamley}}, \bibinfo {author} {\bibfnamefont {C.~S.}\ \bibnamefont {Gerving}}, \bibinfo {author} {\bibfnamefont {T.~M.}\ \bibnamefont {Hoang}}, \bibinfo {author} {\bibfnamefont {E.~M.}\ \bibnamefont {Bookjans}}, \ and\ \bibinfo {author} {\bibfnamefont {M.~S.}\ \bibnamefont {Chapman}},\ }\href {\doibase 10.1038/nphys2245} {\bibfield  {journal} {\bibinfo  {journal} {Nature Physics}\ }\textbf {\bibinfo {volume} {8}},\ \bibinfo {pages} {305} (\bibinfo {year} {2012})}\BibitemShut {NoStop}%
\bibitem [{\citenamefont {Kitzinger}\ \emph {et~al.}(2021)\citenamefont {Kitzinger}, \citenamefont {Meng}, \citenamefont {Fadel}, \citenamefont {Ivannikov}, \citenamefont {Nemoto}, \citenamefont {Munro},\ and\ \citenamefont {Byrnes}}]{KitzingerPRA21}%
  \BibitemOpen
  \bibfield  {author} {\bibinfo {author} {\bibfnamefont {J.}~\bibnamefont {Kitzinger}}, \bibinfo {author} {\bibfnamefont {X.}~\bibnamefont {Meng}}, \bibinfo {author} {\bibfnamefont {M.}~\bibnamefont {Fadel}}, \bibinfo {author} {\bibfnamefont {V.}~\bibnamefont {Ivannikov}}, \bibinfo {author} {\bibfnamefont {K.}~\bibnamefont {Nemoto}}, \bibinfo {author} {\bibfnamefont {W.~J.}\ \bibnamefont {Munro}}, \ and\ \bibinfo {author} {\bibfnamefont {T.}~\bibnamefont {Byrnes}},\ }\href {\doibase 10.1103/PhysRevA.104.043323} {\bibfield  {journal} {\bibinfo  {journal} {Phys. Rev. A}\ }\textbf {\bibinfo {volume} {104}},\ \bibinfo {pages} {043323} (\bibinfo {year} {2021})}\BibitemShut {NoStop}%
\bibitem [{\citenamefont {Luo}\ \emph {et~al.}(2017)\citenamefont {Luo}, \citenamefont {Zou}, \citenamefont {Wu}, \citenamefont {Liu}, \citenamefont {Han}, \citenamefont {Tey},\ and\ \citenamefont {You}}]{OurFriends}%
  \BibitemOpen
  \bibfield  {author} {\bibinfo {author} {\bibfnamefont {X.-Y.}\ \bibnamefont {Luo}}, \bibinfo {author} {\bibfnamefont {Y.-Q.}\ \bibnamefont {Zou}}, \bibinfo {author} {\bibfnamefont {L.-N.}\ \bibnamefont {Wu}}, \bibinfo {author} {\bibfnamefont {Q.}~\bibnamefont {Liu}}, \bibinfo {author} {\bibfnamefont {M.-F.}\ \bibnamefont {Han}}, \bibinfo {author} {\bibfnamefont {M.~K.}\ \bibnamefont {Tey}}, \ and\ \bibinfo {author} {\bibfnamefont {L.}~\bibnamefont {You}},\ }\href {\doibase 10.1126/science.aag1106} {\bibfield  {journal} {\bibinfo  {journal} {Science}\ }\textbf {\bibinfo {volume} {355}},\ \bibinfo {pages} {620} (\bibinfo {year} {2017})}\BibitemShut {NoStop}%
\bibitem [{\citenamefont {Fine}(1982)}]{FinePRL1982}%
  \BibitemOpen
  \bibfield  {author} {\bibinfo {author} {\bibfnamefont {A.}~\bibnamefont {Fine}},\ }\href {\doibase 10.1103/physrevlett.48.291} {\bibfield  {journal} {\bibinfo  {journal} {Physical Review Letters}\ }\textbf {\bibinfo {volume} {48}},\ \bibinfo {pages} {291} (\bibinfo {year} {1982})}\BibitemShut {NoStop}%
\bibitem [{\citenamefont {Chazelle}(1993)}]{ChazelleDCG1993}%
  \BibitemOpen
  \bibfield  {author} {\bibinfo {author} {\bibfnamefont {B.}~\bibnamefont {Chazelle}},\ }\href {\doibase 10.1007/bf02573985} {\bibfield  {journal} {\bibinfo  {journal} {Discrete {\&} Computational Geometry}\ }\textbf {\bibinfo {volume} {10}},\ \bibinfo {pages} {377} (\bibinfo {year} {1993})}\BibitemShut {NoStop}%
\bibitem [{\citenamefont {Pitowsky}\ and\ \citenamefont {Svozil}(2001)}]{PitowskyPRA2001}%
  \BibitemOpen
  \bibfield  {author} {\bibinfo {author} {\bibfnamefont {I.}~\bibnamefont {Pitowsky}}\ and\ \bibinfo {author} {\bibfnamefont {K.}~\bibnamefont {Svozil}},\ }\href {\doibase 10.1103/physreva.64.014102} {\bibfield  {journal} {\bibinfo  {journal} {Physical Review A}\ }\textbf {\bibinfo {volume} {64}} (\bibinfo {year} {2001}),\ 10.1103/physreva.64.014102}\BibitemShut {NoStop}%
\bibitem [{\citenamefont {Tura}\ \emph {et~al.}(2015)\citenamefont {Tura}, \citenamefont {Augusiak}, \citenamefont {Sainz}, \citenamefont {L{\"u}cke}, \citenamefont {Klempt}, \citenamefont {Lewenstein},\ and\ \citenamefont {Ac{\'i}n}}]{AnnPhys}%
  \BibitemOpen
  \bibfield  {author} {\bibinfo {author} {\bibfnamefont {J.}~\bibnamefont {Tura}}, \bibinfo {author} {\bibfnamefont {R.}~\bibnamefont {Augusiak}}, \bibinfo {author} {\bibfnamefont {A.}~\bibnamefont {Sainz}}, \bibinfo {author} {\bibfnamefont {B.}~\bibnamefont {L{\"u}cke}}, \bibinfo {author} {\bibfnamefont {C.}~\bibnamefont {Klempt}}, \bibinfo {author} {\bibfnamefont {M.}~\bibnamefont {Lewenstein}}, \ and\ \bibinfo {author} {\bibfnamefont {A.}~\bibnamefont {Ac{\'i}n}},\ }\href {\doibase http://doi.org/10.1016/j.aop.2015.07.021} {\bibfield  {journal} {\bibinfo  {journal} {Annals of Physics}\ }\textbf {\bibinfo {volume} {362}},\ \bibinfo {pages} {370 } (\bibinfo {year} {2015})}\BibitemShut {NoStop}%
\bibitem [{\citenamefont {Fukuda}(1997)}]{fukuda1997cdd}%
  \BibitemOpen
  \bibfield  {author} {\bibinfo {author} {\bibfnamefont {K.}~\bibnamefont {Fukuda}},\ }\href@noop {} {\bibfield  {journal} {\bibinfo  {journal} {Institute for Operations Research, ETH-Zentrum}\ ,\ \bibinfo {pages} {91}} (\bibinfo {year} {1997})}\BibitemShut {NoStop}%
\bibitem [{\citenamefont {Aloy}\ \emph {et~al.}()\citenamefont {Aloy}, \citenamefont {M\"uller-Rigat}, \citenamefont {Lewenstein}, \citenamefont {Tura},\ and\ \citenamefont {Fadel}}]{UsInPrep}%
  \BibitemOpen
  \bibfield  {author} {\bibinfo {author} {\bibfnamefont {A.}~\bibnamefont {Aloy}}, \bibinfo {author} {\bibfnamefont {G.}~\bibnamefont {M\"uller-Rigat}}, \bibinfo {author} {\bibfnamefont {M.}~\bibnamefont {Lewenstein}}, \bibinfo {author} {\bibfnamefont {J.}~\bibnamefont {Tura}}, \ and\ \bibinfo {author} {\bibfnamefont {M.}~\bibnamefont {Fadel}},\ }\href@noop {} {\enquote {\bibinfo {title} {Bell inequalities as a tool to probe quantum chaos},}\ }\Eprint {http://arxiv.org/abs/arXiv today} {arXiv today} \BibitemShut {NoStop}%
\bibitem [{\citenamefont {M\"uller-Rigat}\ \emph {et~al.}()\citenamefont {M\"uller-Rigat}, \citenamefont {Aloy}, \citenamefont {Lewenstein}, \citenamefont {Fadel},\ and\ \citenamefont {Tura}}]{UsInPrep2}%
  \BibitemOpen
  \bibfield  {author} {\bibinfo {author} {\bibfnamefont {G.}~\bibnamefont {M\"uller-Rigat}}, \bibinfo {author} {\bibfnamefont {A.}~\bibnamefont {Aloy}}, \bibinfo {author} {\bibfnamefont {M.}~\bibnamefont {Lewenstein}}, \bibinfo {author} {\bibfnamefont {M.}~\bibnamefont {Fadel}}, \ and\ \bibinfo {author} {\bibfnamefont {J.}~\bibnamefont {Tura}},\ }\href@noop {} {\enquote {\bibinfo {title} {Three-outcome multipartite bell inequalities: applications to dimension witnessing and spin-nematic squeezing in many-body systems},}\ }\Eprint {http://arxiv.org/abs/arXiv today} {arXiv today} \BibitemShut {NoStop}%
\bibitem [{\citenamefont {Aloy}\ \emph {et~al.}(2021)\citenamefont {Aloy}, \citenamefont {Fadel},\ and\ \citenamefont {Tura}}]{AloyQMP}%
  \BibitemOpen
  \bibfield  {author} {\bibinfo {author} {\bibfnamefont {A.}~\bibnamefont {Aloy}}, \bibinfo {author} {\bibfnamefont {M.}~\bibnamefont {Fadel}}, \ and\ \bibinfo {author} {\bibfnamefont {J.}~\bibnamefont {Tura}},\ }\href {\doibase 10.1088/1367-2630/abe15e} {\bibfield  {journal} {\bibinfo  {journal} {New Journal of Physics}\ }\textbf {\bibinfo {volume} {23}},\ \bibinfo {pages} {033026} (\bibinfo {year} {2021})}\BibitemShut {NoStop}%
\bibitem [{\citenamefont {Fadel}\ and\ \citenamefont {Tura}(2017)}]{FadelPRL2017}%
  \BibitemOpen
  \bibfield  {author} {\bibinfo {author} {\bibfnamefont {M.}~\bibnamefont {Fadel}}\ and\ \bibinfo {author} {\bibfnamefont {J.}~\bibnamefont {Tura}},\ }\href {\doibase 10.1103/PhysRevLett.119.230402} {\bibfield  {journal} {\bibinfo  {journal} {Phys. Rev. Lett.}\ }\textbf {\bibinfo {volume} {119}},\ \bibinfo {pages} {230402} (\bibinfo {year} {2017})}\BibitemShut {NoStop}%
\bibitem [{\citenamefont {Grant}\ and\ \citenamefont {Boyd}(2014)}]{CVX1}%
  \BibitemOpen
  \bibfield  {author} {\bibinfo {author} {\bibfnamefont {M.}~\bibnamefont {Grant}}\ and\ \bibinfo {author} {\bibfnamefont {S.}~\bibnamefont {Boyd}},\ }\href@noop {} {\enquote {\bibinfo {title} {{CVX}: Matlab software for disciplined convex programming, version 2.1},}\ }\bibinfo {howpublished} {\url{http://cvxr.com/cvx}} (\bibinfo {year} {2014})\BibitemShut {NoStop}%
\bibitem [{\citenamefont {Skrzypczyk}\ and\ \citenamefont {Cavalcanti}(2023)}]{BookSDP}%
  \BibitemOpen
  \bibfield  {author} {\bibinfo {author} {\bibfnamefont {P.}~\bibnamefont {Skrzypczyk}}\ and\ \bibinfo {author} {\bibfnamefont {D.}~\bibnamefont {Cavalcanti}},\ }\href {\doibase 10.1088/978-0-7503-3343-6} {\emph {\bibinfo {title} {Semidefinite Programming in Quantum Information Science}}},\ 2053-2563\ (\bibinfo  {publisher} {IOP Publishing},\ \bibinfo {year} {2023})\BibitemShut {NoStop}%
\bibitem [{\citenamefont {Lasserre}(2009)}]{Lasserre2009}%
  \BibitemOpen
  \bibfield  {author} {\bibinfo {author} {\bibfnamefont {J.~B.}\ \bibnamefont {Lasserre}},\ }\href {https://www.ebook.de/de/product/27808945/jean_bernard_lasserre_moments_positive_polynomials_and_their_applications.html} {\emph {\bibinfo {title} {Moments, Positive Polynomials and Their Applications}}}\ (\bibinfo  {publisher} {IMPERIAL COLLEGE PR},\ \bibinfo {year} {2009})\BibitemShut {NoStop}%
\bibitem [{\citenamefont {Gouveia}\ \emph {et~al.}(2010)\citenamefont {Gouveia}, \citenamefont {Parrilo},\ and\ \citenamefont {Thomas}}]{Gouveia2010}%
  \BibitemOpen
  \bibfield  {author} {\bibinfo {author} {\bibfnamefont {J.}~\bibnamefont {Gouveia}}, \bibinfo {author} {\bibfnamefont {P.~A.}\ \bibnamefont {Parrilo}}, \ and\ \bibinfo {author} {\bibfnamefont {R.~R.}\ \bibnamefont {Thomas}},\ }\href {\doibase 10.1137/090746525} {\bibfield  {journal} {\bibinfo  {journal} {{SIAM} Journal on Optimization}\ }\textbf {\bibinfo {volume} {20}},\ \bibinfo {pages} {2097} (\bibinfo {year} {2010})}\BibitemShut {NoStop}%
\bibitem [{\citenamefont {Gouveia}\ and\ \citenamefont {Thomas}(2012)}]{Gouveia2012}%
  \BibitemOpen
  \bibfield  {author} {\bibinfo {author} {\bibfnamefont {J.}~\bibnamefont {Gouveia}}\ and\ \bibinfo {author} {\bibfnamefont {R.~R.}\ \bibnamefont {Thomas}},\ }in\ \href {\doibase 10.1137/1.9781611972290.ch7} {\emph {\bibinfo {booktitle} {Semidefinite Optimization and Convex Algebraic Geometry}}}\ (\bibinfo  {publisher} {Society for Industrial and Applied Mathematics},\ \bibinfo {year} {2012})\ pp.\ \bibinfo {pages} {293--340}\BibitemShut {NoStop}%
\end{thebibliography}%

\end{document}